\documentclass[format=acmsmall, screen=true, review=false]{acmart}
\usepackage{booktabs}
\usepackage{tikz}
\usetikzlibrary{external}
\usepackage{ textcomp }
\usepackage[utf8]{inputenc}
\usepackage{enumitem}
\newcommand{\xsub}[1]{%
  \mbox{\scriptsize\begin{tabular}{@{}c@{}}#1\end{tabular}}%
}

\usepackage{makecell}
\usepackage{threeparttable}
\usepackage{longtable}
\usepackage{multirow}
\usepackage{tabularx}
\usepackage{tabulary}
\usepackage{tablefootnote}
\usepackage{array}
\usepackage{hyperref}
\newcommand{\PreserveBackslash}[1]{\let\temp=\\#1\let\\=\temp}

\newcolumntype{C}[1]{>{\PreserveBackslash\centering}p{#1}}
\newcolumntype{R}[1]{>{\PreserveBackslash\raggedleft}p{#1}}
\newcolumntype{L}[1]{>{\PreserveBackslash\raggedright}p{#1}}
\usepackage{amssymb}
\usepackage{amsmath}
\usepackage{amsthm}
\usepackage{epstopdf}
\usepackage{graphicx}
\usepackage{subfig}
\usepackage{caption}
\usepackage{color}
\usepackage{url}
\usepackage{hyperref}
\usepackage{bm}
\usepackage{algorithmic}
\usepackage[linesnumbered,ruled,vlined]{algorithm2e}

\SetCommentSty{mycommfont}

\usepackage[ruled]{algorithm2e} 

\SetAlFnt{\small}
\SetAlCapFnt{\small}
\SetAlCapNameFnt{\small}
\SetAlCapHSkip{0pt}
\IncMargin{-\parindent}

\acmJournal{TOIS}
\acmVolume{x}
\acmNumber{x}
\acmArticle{1}
\acmYear{2018}
\acmMonth{xx}
\copyrightyear{2018}

\setcopyright{acmlicensed}

\acmDOI{0000001.0000001}

\begin{document}

\title{MMALFM: Explainable Recommendation by Leveraging Reviews and Images}

\author{Zhiyong Cheng}
 \affiliation{%
  \institution{ Qilu University of Technology (Shandong Academy of Sciences)}
  \streetaddress{ Shandong Computer Science Center (National Supercomputer Center in Jinan), Shandong Artificial Intelligence Institute}
  \country{China}
 }
\email{jason.zy.cheng@gmail.com}
\author{Xiaojun Chang}
\affiliation{%
 \institution{Monash University}
 \streetaddress{Information Technology}
 \country{Australia}}
\email{cxj273@gmail.com}
\author{Lei Zhu}
\affiliation{
    \institution{Shandong Normal University}
    \streetaddress{College of Computer Science and Electronic Engineering}
    \city{Jinan}
    \state{Shandong}
    \postcode{250358}
    \country{China}
}
\email{leizhu0608@gmail.com }

\author{Rose C. Kanjirathinkal}
\affiliation{%
  \institution{Carnegie Mellon University}
  \streetaddress{5501, Gates-Hillman Center}
   \city{Pittsburgh}
   \state{PA}
   \postcode{15213}
 \country{USA}}
\email{rosecatherinek@cs.cmu.edu}
\author{Mohan Kankanhalli}
\affiliation{%
  \institution{National University of Singapore}
  \streetaddress{13 Computing Drive}
  \postcode{117417}
  \country{Singapore}
}
\email{mohan@comp.nus.edu.sg}
\renewcommand{\shortauthors}{Z. Cheng et al.}

\begin{abstract}
Personalized rating prediction is an important research problem in recommender systems. Although the latent factor model (e.g., matrix factorization) achieves good accuracy in rating prediction, it suffers from many problems including cold-start, non-transparency, and suboptimal results for individual user-item pairs. In this paper, we exploit textual reviews and item images together with ratings to tackle these limitations.  Specifically, we first apply a proposed multi-modal aspect-aware topic model (MATM) on text reviews and item images to model users' preferences and items' features from different \emph{aspects},  and also estimate the \emph{aspect importance} of a user towards an item. Then the aspect importance is  integrated into a novel aspect-aware latent factor model (ALFM), which learns user's and item's latent factors based on ratings. In particular, ALFM introduces a weight matrix to associate those latent factors with the same set of aspects in MATM, such that the latent factors could be used to estimate aspect ratings. Finally, the overall rating is computed via a linear combination of the aspect ratings, which are weighted by the corresponding aspect importance.  To this end, our model could alleviate the data sparsity problem and gain good interpretability for recommendation. Besides, every aspect rating is weighted by its aspect importance, which is dependent on the targeted user's preferences and the targeted item's features. Therefore, it is expected that the proposed method can model a user's preferences on an item more accurately for each user-item pair. Comprehensive experimental studies have been conducted on the Yelp 2017 Challenge dataset and Amazon product datasets.  Results show that (1) our method achieves significant improvement compared to strong baseline methods, especially for users with only few ratings; (2) item visual features can improve the prediction performance - the effects of item image features on improving the prediction results depend on the importance of the visual features for the items; and (3) our model can explicitly interpret the predicted results in great detail.

\end{abstract}

\setcopyright{acmcopyright}
\acmJournal{TOIS}
\acmYear{2019} \acmVolume{1} \acmNumber{1} \acmArticle{1} \acmMonth{1} \acmPrice{15.00}\acmDOI{10.1145/3291060}



\maketitle

\section{Introduction}
Nowadays, referring the ratings of targeted products in the online review/E-commerce websites, such as Yelp\footnote{https://www.yelp.com/} and Amazon\footnote{https://www.amazon.com/}, becomes a nature behavior for users to make decisions in daily consumption. A product rating reflects a user's overall satisfaction or judgment on the product.  Accordingly, predicting the ratings for products is a practical way to increase revenue for the E-commerce websites, as it could guide the recommendation of products to
potential customers. In fact, \emph{personalized rating prediction} has been raised as an important research problem in recommender systems since the Netflix Prize contest~\cite{bell2007lessons}. As demonstrated in the contest, latent factor models (e.g., matrix factorization~\cite{koren2008factorization,koren2009matrix}) are the most widely used and successful techniques for rating prediction. These methods characterize user's interests and item's features using \emph{latent factors} inferred from rating patterns in user-item rating records.  However, as a typical collaborative filtering technique, the MF-based method easily suffers from the \emph{cold-start} problem - when there are only few ratings for items or users~\cite{he2015trirank,ren2017social}, the performance deteriorates dramatically. Besides, a rating only indicates the overall satisfaction of a user towards an item, it cannot explain the underlying rationale properly. For example, a user could give a restaurant a high rating because of its delicious food or due to its nice ambience. Most existing MF models cannot provide such fine-grained analysis.  Therefore, relying solely on ratings makes it hard for these methods to explicitly and accurately model user's preferences~\cite{he2015trirank,ling2014ratings,mcauley2013hidden,wu2015flame}.

The above two limitations have been widely discussed and studied. For example, various types of side information have been incorporated into MF to alleviate the cold-start problem, such as tags~\cite{shi2013mining}, social relations~\cite{ma2011recommender,ren2017social}, reviews~\cite{ling2014ratings,mcauley2013hidden,zhang2016integrating}, visual features~\cite{he2016vbpr}, and contextual information~\cite{cheng2014just,cheng2016effective}. Among them, the accompanying review of a rating contains important complementary information, which not only encodes the information about user preferences and item features but also explains the underlying reasons for the rating. Therefore, in recent years, many models have been developed to exploit reviews with ratings to tackle the cold-start problem and also enhance the explainability of MF, such as  HFT~\cite{mcauley2013hidden}, CTR~\cite{wang2011collaborative},  RMR~\cite{ling2014ratings}, RBLT~\cite{tan2016rating},  and ITLFM~\cite{zhang2016integrating}. However, a limitation of these models is that they all assume a \emph{one-to-one correspondence relationship} between latent topics (learned from reviews) and latent factors (learned from ratings), which not only limits their flexibility on modeling reviews and ratings but also may not be optimal.

While substantial progress has been achieved so far, another limitation of matrix factorization model has seldom been  discussed - \emph{each latent factor in MF is treated uniformly, which may result in sub-optimal recommendation results}. 
In the training stage, MF learns the latent factors of users ($\bm{p_u}$) and items ($\bm{q_i}$) via a global optimization strategy~\cite{christakopoulou2016local}. In other words,  $\bm{p_u}$ and $\bm{q_i}$ are optimized to achieve a global optimization over all the user-item ratings in the training dataset.\footnote{In the paper, unless otherwise specified, notations in bold style denote matrices or vectors, and the ones in normal style denote scalars.} As a result, the performance could be severely compromised locally for individual users or items. In the testing stage, an unknown rating is predicted by the dot product of the targeted user $u$'s and item $i$'s latent factors, namely, the linear combination of $p_{u,k}*q_{i,k}$ for each factor $k$ with the same weight.
However, the relative importance of a factor with respect to different user-item pairs could be very different. For example, a user expects \emph{high-quality service} and \emph{decent ambience} for an expensive restaurant; while for a cheap restaurant, the expectation on these two aspects would be low. Thus, the user will give higher weights to the aspects of ``service" and ``ambience" for the expensive restaurant than the cheap one when rating two such restaurants. Therefore, for accurate prediction, it is important to accurately capture the importance of each latent factor for a user towards an item. At first glance, MF achieves the goal as the influence of a factor (e.g., $k$-th factor) is dependent on both $p_{u,k}$ and $q_{i,k}$ (i.e., $p_{u,k}*q_{i,k}$).  However, it models the importance of a factor by a fixed value for an item or a user.  As a result, it treats each factor of an item with the same importance to all users (i.e., $q_{i,k}$); and similarly, each factor of a user is equally important to all items (i.e., $p_{u,k}$) in rating prediction. This will lead to sub-optimal results for individual user-item pair. 

In this work, we attempt to address the limitations of \emph{cold-start}, \emph{non-transparency}, and \emph{sub-optimal results} in MF  simultaneously. Specifically, we associate the latent factors in MF to the explainable ``aspects" in textual reviews and item images. Review comments complement ratings by providing preferences of users and notable features of items in different aspects. For a specific type of item, the \emph{aspects} that users care about  can be easily observed, such as \emph{food}, \emph{service}, \emph{ambience},  and \emph{price} for \emph{restaurants}. Different users may care about different aspects of an item. For example,  some users care about the taste of food while some others pay attention to the ambience of restaurants. Besides, even for the same aspect,  the preference of users could be different from each other. For instance, some users like \emph{Chinese cuisine} while some others favor \emph{Italian cuisine}. In general, users tend to discuss more about the aspects they are more interested in. Therefore, the evidence in the reviews could be leveraged to estimate user's attention on different aspects. The overall rating of a user towards an item is highly dependent on the user's satisfaction on different aspects (i.e., \emph{aspect ratings}) and the importance of those aspects for the user with respect to the item (i.e., \emph{aspect importance}).  To accurately predict the overall rating, it is important to capture users' preference and items' characteristics in different aspects by analyzing and modeling user's reviews and ratings, so as to estimate the aspect ratings and aspect importance. Besides, for some types of items (e.g., \emph{clothing}), their visual appearances play an important role in their properties, which can greatly bias users' preferences towards them~\cite{he2016vbpr,he2016vista,mcauley2015image}. For example, users can easily determine whether they like a restaurant based on the images of \emph{food} and \emph{interior ambience} of the restaurant.  Thus, the visual features of items are also important complementary information for modeling items' characteristics.
\begin{figure}
    \centering
        \includegraphics[width = 14cm]{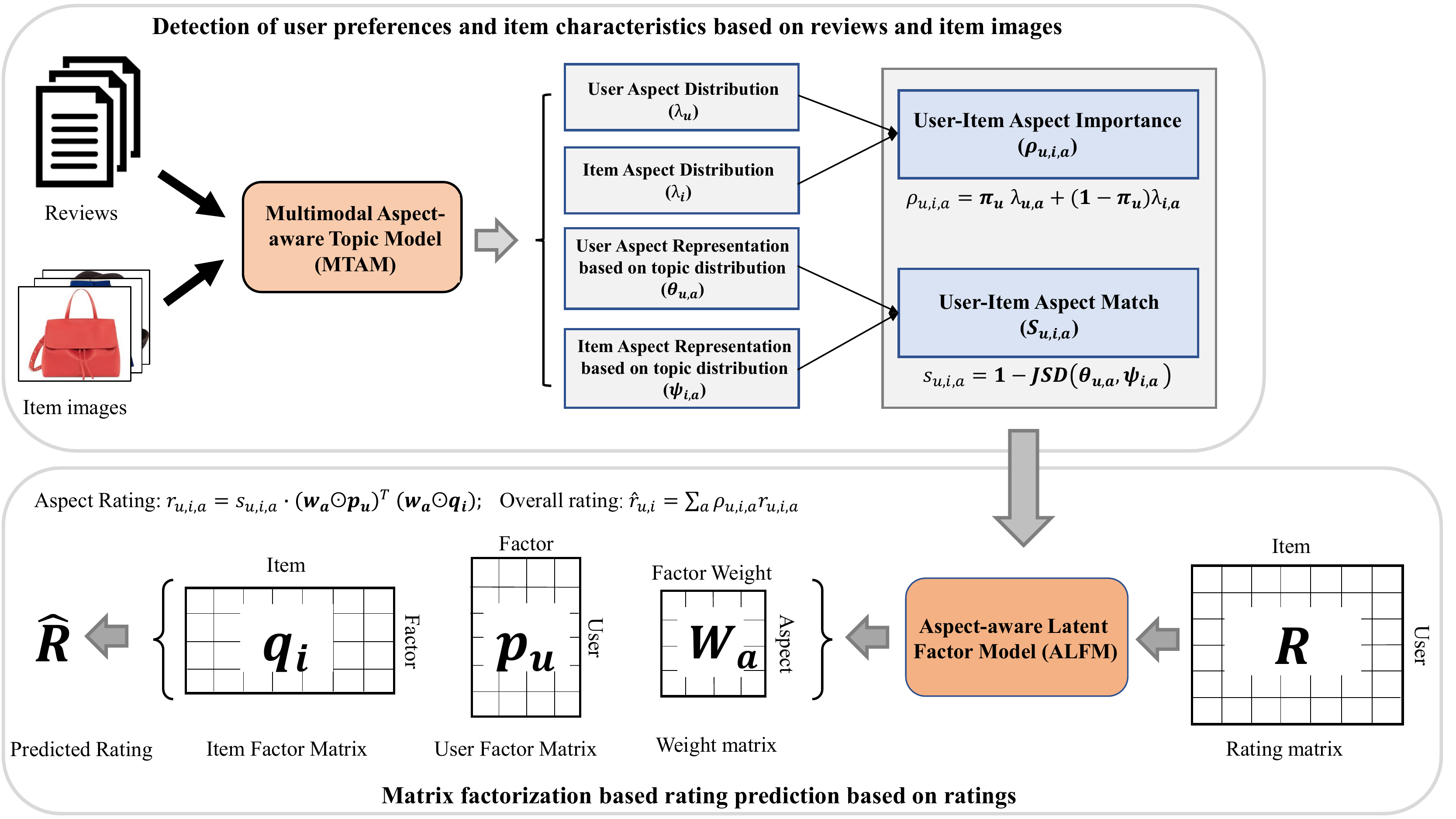}
   \caption{An overview of our proposed recommender system. }
       \label{fig:overview}
\end{figure}

We propose  a multi-modal aspect-aware topic model to utilize user reviews and item images together to learn shared latent topics, which are used to model users' preferences and items' properties in different aspects, as well as to estimate the importance of an aspect (i.e., \emph{aspect importance}) for a user towards an item.
The results are then integrated into a proposed aspect-aware latent factor model (ALFM) to estimate the \emph{aspect ratings} for the final overall rating prediction. In particular, a weight matrix is introduced in ALFM to associate the latent factors to different aspects, such that the model is able to predict aspect ratings.  In this way, our model avoids referring to external sentiment analysis tools for aspect rating prediction as in~\cite{zhang2014explicit,diao2014jointly}. The overall rating is obtained by a linear combination of the \emph{aspect ratings}, which are weighted by the importance of corresponding aspects (i.e., \emph{aspect importance}). An overview of the proposed remmender system is shown in Fig.~\ref{fig:overview}.  Note that in our method, both the latent topics and latent factors are used to represent the same set of aspects. Therefore, the latent topics and the latent factors are thus correlated on the ``aspect" level. This is fundamentally different from previous review-based rating prediction models, which assume a \emph{one-to-one correspondence relationship} between the latent topics (learned from reviews) and the latent factors (learned from rating), such as the models in~\cite{bao2014topicmf,ling2014ratings,mcauley2013hidden,zhang2016integrating}. Besides, our model could learn an aspect importance for each user-item pair, namely, assigning a different weight to each latent factor $p_{u,k}*q_{i,k}$\footnote{Details can be found in Eq.~\ref{eq:re2}}, and thus could alleviate the sub-optimal local recommendation problem and achieve better performance.

Note that we presented a preliminary study in~\cite{cheng2017aspect}. In this paper, we substantially extended~\cite{cheng2017aspect} from multiple perspectives. Firstly, we additionally considered item images to model users' preferences and items' features from different aspects, which has not been studied in the previous work. Secondly, due to the consideration of images, we designed a new multi-modal topic model for user preference modeling and specified the detailed inference procedure of the topic model. Besides, we also added  discussions about the difference of our model with previous recommendation models, which consider both reviews and ratings. Finally, we conducted extensively new experiments on the public Yelp 2017 and Amazon product datasets. In addition, we added more competitors and also evaluated the performance of our model on the top-n recommendation task by standard ranking metrics such as precision and NDCG.  We detailedly studied the effects of item visual features and find that not all visual features are useful in rating prediction.

To sum up, our primary contributions are listed as follows:

\begin{itemize} [align=left,style=nextline,leftmargin=*,labelsep=\parindent,font=\normalfont]
\item We propose a novel aspect-aware latent factor model. Our model firstly utilizes both textual reviews and image features to learn users' preferences on different aspects, and then integrates the learned preferences into rating-based matrix factorization model for accurate rating prediction. Particularly, our model relaxes the constraint of one-to-one mappings between the latent topics and latent factors in previous models. Thus, our model is flexible to tune the two parameters separately for better performance.

\item Our model can automatically learn the aspect importance/weights for different user-item pairs. By associating latent factors with aspects, the aspect weights are integrated with latent factors  for rating prediction. Thus, the proposed model could alleviate the sub-optimal recommendation results in MF for individual user-item pairs.

\item We conduct comprehensive experimental studies on several benchmark datasets to evaluate the effectiveness of our model. Results show that our model is significantly better than previous approaches on tasks of rating prediction,  recommendation for sparse data, and recommendation interpretability.
\end{itemize}

The remainder of this paper is organized as follows: Section~\ref{sec:relwork} gives a brief overview of related work; Section~\ref{sec:ourmodel} detailedly describes the proposed model including a multimedia aspect-aware topic model and an aspect-aware latent factor model as well as their inference; Section~\ref{sec:expconfig} describes the experimental datasets and configurations; and Section~\ref{sec:expresult} presents and analyzes the experimental results. Finally, Section~\ref{sec:concl} concludes the paper.


\section{Related Work} \label{sec:relwork}
A comprehensive review on recommender systems is beyond the scope of this work. We mainly discuss closely related works in the following three categories: (1) review-aware recommendation; (2) visually-aware recommendation; and (3) multi-modal topic model.

\subsection{Review-aware Recommendation}
 We mainly discuss the works which utilize both reviews and ratings for rating prediction. Some works assume that the review is available when predicting the rating score, such as SUIT~\cite{li2014suit}, LARAM~\cite{wang2011latent}, and recent DeepCoNN~\cite{zheng2017joint}.  However, in real world recommendation settings, the task should be predicting ratings for the uncommented and unrated items. Therefore, the review is unavailable when predicting ratings. In recent years, many works have been proposed to combine reviews and ratings to improve the rating prediction performance for this scenario. We broadly classify the approaches for this task in three categories: (1) sentiment-based, (2) topic-based, and (3) deep learning-based. Our approach falls into the second category.

\textbf{Sentiment-based.} These works analyze user's sentiments on items in reviews to boost the rating prediction performance, such as~\cite{pappas2013sentiment,pero2013opinion,diao2014jointly,zhang2014explicit, bauman2017aspect}. For example, ~\cite{pappas2013sentiment} estimates a sentiment score for each review to build a user-item sentiment matrix, then a traditional collaborative filtering method is applied. Zhang et al.~\cite{zhang2014explicit} analyze the sentiment polarities of reviews and then jointly factorize the user–item rating matrix. Bauman et al.~\cite{bauman2017aspect} presented a SULM model which extracts aspects and  classifies sentiments on the aspects in user reviews, aiming to recommend item together with the most important aspects that may enhance user experience. These methods rely on the performance of external NLP tools for sentiment analysis and thus are not self-contained.

\textbf{Topic-based.} These approaches extract latent topics or aspects from reviews. An early work~\cite{ganu2009beyond} in this direction relies on domain knowledge to manually label reviews into different aspects, which requires expensive domain knowledge and high labor cost. Later on, most works attempt to extract latent topics or aspects from reviews automatically~\cite{mcauley2013hidden,bao2014topicmf,diao2014jointly,he2015trirank,ling2014ratings,mcauley2013hidden,wu2015flame,zhang2016integrating,tan2016rating}. A general approach of these methods is to extract latent topics from reviews using topic models~\cite{wang2011collaborative,mcauley2013hidden,ling2014ratings,zhang2016integrating,tan2016rating} or non-negative MF~\cite{bao2014topicmf,qiu2016aspect} and learn latent factors from ratings using MF methods. HFT~\cite{mcauley2013hidden} and  TopicMF~\cite{bao2014topicmf} link the latent topics and latent factors by using a defined transform function.    ITLFM~\cite{zhang2016integrating} and RBLT~\cite{tan2016rating} assume that the latent topics and latent factors are in the same space, and linearly combine them to form the latent representations for users and items to model the ratings in MF. CTR~\cite{wang2011collaborative} assumes that the latent factors of items depend on the latent topic distributions of their text, and adds a latent variable to offset the topic distributions of items when modeling the ratings. RMR~\cite{ling2014ratings} also learns item's features using topic models on reviews, while it models ratings using a mixture of Gaussian rather than MF methods. Diao et al.~\cite{diao2014jointly} propose an integrated graphical model called JMARS to jointly model aspects, ratings and sentiments for movie rating prediction. Those models all assume an one-to-one mapping between the learned latent topics from reviews and latent factors from ratings.  Although we adopt the same strategy to extract latent topics and learn latent factors, our model does not have the constraint of one-to-one mapping.\footnote{Notice that in our model, an aspect is represented as a distribution of latent topics, and thus a latent topic can be regarded as  a subaspect. In most of the previous works, the latent topics or latent factors learned from ratings or reviews are regarded as an aspect, which is quite different.} Besides, Zhang et al.~\cite{zhang2014explicit} extract aspects by decomposing the user–item rating matrix into item–aspect and user–aspect matrices.   He et al.~\cite{he2015trirank}  extract latent topics from reviews by modeling the user-item-aspect relation with a tripartite graph.

\textbf{Deep learning-based}. A recent research trend is to leverage deep learning for recommendation. For example, ~\cite{li2015deep,wu2016collaborative} use auto-encoder approach for top-n recommendation. He et al.~\cite{he2017neural,he2017facorization} generalize matrix factorization and factorization machines for neural collaborative filtering. Researchers also attempt to apply deep textual modeling on reviews for recommendation~
\cite{zhang2016collaborative,zheng2017joint,cheng2018ancf,catherine2017transnets,zhang2017joint,chen2018neural}. ~\cite{zhang2017joint} apply a multi-modal deep learning framework to fuse heterogeneous information sources for top-n recommendation.  In DeepCoNN~\cite{zheng2017joint}, the concatenation of all reviews of a user or an item is used as an input to a CNN to learn a representation of the user or the item. Then representations of users and items are concatenated and passed into a regression layer for rating prediction. Although experimental results show that DeepCoNN outperforms CTR~\cite{wang2011collaborative} and HFT~\cite{mcauley2013hidden}, DeepCoNN uses reviews in the testing phase. ~\cite{catherine2017transnets} shows that the performance of DeepCoNN decreases greatly when reviews are unavailable in the testing phase. To avoid using reviews in rating prediction,  ~\cite{catherine2017transnets} develops a TransNet, which extends the DeepCoNN by introducing an additional layer to represent an approximation of the review corresponding to the target user-item pair.  However, authors did not compare TransNet with other strong baselines, such as HFT and CTR. A recent work proposed by Cheng et al.~\cite{cheng2018ancf} shares the same spirit with this work: considering users' varied attentions on different aspects of items.  Different from the method in this work based on matrix factorization, they proposed an A$^3$NCF model which adopts an attentive neural network to capture users' attentions on different aspects.

\subsection{Visually-aware Recommendation}
With the advancement of techniques on image analysis, researchers have started to pay attentions to the visual appearances of items in recommender systems, especially for the items of which the visual appearances are crucial side-signals, such as \emph{clothing}~\cite{mcauley2015image,he2016vbpr,he2016vista}. For example, ~\cite{jagadeesh2014large} and ~\cite{kalantidis2013getting} consider visual features specifically for the task of clothing recommendation. Their methods are specially designed for clothing recommendation and require handcrafted methods and carefully annotated data. McAuley et al.~\cite{mcauley2015image} develop a recommender system to recommend clothes and accessories by modeling users' visual preferences with the use of visual contents extracted from cloth and accessory images.   He et al.~\cite{he2016sherlock,he2016vbpr} extend the standard MF with visual dimensions to facilitate the fashion item recommendation. Long-term temporal dynamics (e.g., fashion evolution) and \emph{``visual consistently"} in session-level user actions are considered in their following works~\cite{he2016ups,he2016vista}. Chen et al.~\cite{chen2016context} utilize visual features for personalized image tweet recommendation. Wang et al.~\cite{wang2017your} explore images for POI recommendation by incorporating visual contents in PMF~\cite{salakhutdinov2007probabilistic}. More recently, ~\cite{zhang2017joint} integrate images with reviews and ratings in a multimodal deep learning framework for top-n recommendation. Most of the above works target the visually-aware personalized ranking problems and rely on visual interactions to incorporate user visual preferences in MF methods (e.g., PMF~\cite{salakhutdinov2007probabilistic} and BPR~\cite{rendle2009bpr}). Different from them, in this work, item visual features are used together with text reviews to model aspect-aware users' preferences and items' properties for rating prediction.

\subsection{Multi-modal Topic Model}
Topic models, such as LDA~\cite{blei2003latent}, have achieved great success in single modality scenarios, and thus they have been extended to support multi-modal case~\cite{barnard2003matching,blei2003modeling,putthividhy2010topic,jia2011learning,virtanen2012factorized,qian2016multi}. The underlying assumption is that there exists shared latent topics which explain the correlations between different modalities. mmLDA~\cite{barnard2003matching} assumes the image and text words are generated from two non-overlapping sets of hidden topics. In Corr-LDA~\cite{blei2003modeling}, image is the primary modality and is generated first, and the caption word is then generated based on the topic of an image region which the word is associated with.
Tr-mmLDA~\cite{putthividhy2010topic} uses a latent variable regression approach to learn a linear mapping between the topic distributions of two modalities. Multi-modal document random field (MDRF)~\cite{jia2011learning} generalizes the modeling of two modalities to multiple modalities. ~\cite{qian2016multi} presents a MMTOM model, which extend mm-LDA to not only learn  multi-modal topics including textual topics and visual topics but also mine opinions of the learned topics in multiple views. In this work, we assume that text words and visual words are generated based on the topic distributions of aspects, which they are assigned to. Cheng et al.~\cite{cheng2016sigir} developed a dual-layer music preference topic model to  construct a shared latent music interest space and characterize the correlations among (user, song, term) by leveraging both audio features and textual tags. In~\cite{cheng2017sigir}, a User-Information-Aware Music Interest Topic (UIA-MIT) model was proposed to model users music preferences by considering audio features, music tag features, as well as user-specific information (e.g., age).

\begin{table}[]
\centering
\small
\caption{Notations and their definitions}
\begin{tabular}{ll}
\Xhline{1.2pt}\noalign{\smallskip}
    Notation & Definition \\
\noalign{\smallskip}\hline\noalign{\smallskip}
        $\mathcal{D}$ & dataset \\
        $\mathcal{P}_i$ & image set of item $i$ \\
        $d_{u,i}$ &  review document of user $u$ to item $i$ \\
        $\mathcal{U}$, $\mathcal{I}$, $\mathcal{A}$ &  user set, item set, and aspect set, respectively \\
        $M,N,A$ & number of users, items, and aspects, respectively\\
        $N_{w,s}$ & number of words in a review $d$ \\
        $N_{v,p}$ & number of visual words in an item image $p$\\
    \noalign{\smallskip}\hline\noalign{\smallskip}
        $K$ &  number of latent topics in MATM  \\
        $y$ & an indicator variable in MATM\\
        $a_s$ & assigned aspect $a$ to a sentence $s$ \\
        $a_p$ & assigned aspect $a$ to an item image $p$ \\
        $t, c$ & a textual and a visual term in the vocabulary, respectively \\
        $w, v$ & a textual word and a visual word in documents \tablefootnote{We adopt the terminology of \emph{term} and \emph{word} in~\cite{heinrich2005}: \emph{Term} refers to the element of a vocabulary, and \emph{word} refers to the element of a document, respectively. A term can be instantiated by several words in a text corpus.} \\
        $T, C$ & the size of textual and visual vocabulary, respectively \\
        $\pi_u$ & the parameter of Bernoulli distribution $P(y=0)$  \\
        $\eta$ & Beta priors ($\eta=\{\eta_0, \eta_1\}$) \\
        $\bm{\alpha_u}, \bm{\alpha_i}$ & Dirichlet priors for aspect-topic distributions\\
        $\bm{\gamma_u}, \bm{\gamma_i} $ & Dirichlet priors for aspect distributions\\
        $\bm{\beta_t}, \bm{\beta_c}  $ & Dirichlet priors for topic-text \& visual word distributions, respectively\\
        $\bm{\theta_{u,a}}$ & user's aspect-topic distribution: denoting user's preference on $a$ \\
        $\bm{\psi_{i,a}}$ & item's aspect-topic distribution: denoting  item's features on $a$\\
        $\bm{\lambda_u}, \bm{\lambda_i}$ & aspect distributions of user and item, respectively \\
        $\bm{\phi_t}, \bm{\phi_c}$  & topic-text and topic-visual word distribution, respectively  \\
    \noalign{\smallskip}\hline\noalign{\smallskip}
        $f$    & number of latent factors in ALFM \\
        $\mu_\cdot$ & regularization coefficients \\
        $b_\cdot$ & bias terms, e.g., $b_u, b_i, b_0$ \\
        $\bm{w_a}$ & weight vector for aspect $a$ \\
        $\bm{p_u}$, $\bm{q_i}$ & latent factors of user $u$ and item $i$, respectively \\
        $r_{u,i}$ & rating of user $u$ to item $i$ \\
        $r_{u,i, a}$ & aspect rating on aspect $a$ of user $u$ towards item $i$ \\
        $\rho_{u,i,a}$ & aspect importance of $a$ for $u$ with respect to $i$ \\
        $s_{u,i,a}$   & denotes the degree of item $i$'s attributes on aspect $a$ matching user $u$'s  \\
                    &  preference on aspect $a$\\
        \noalign{\smallskip}\Xhline{1.2pt}
\end{tabular}
 \label{tab:notation}
\end{table}

\vspace{-2pt}
\section{The Proposed Model} \label{sec:ourmodel}
\subsection{Problem Setting}
Let $\mathcal{D}$ be a collection of reviews of items $\mathcal{I}$ from a specific category (e.g., restaurant) by a set of users $\mathcal{U}$, and each review comes with an overall rating $r_{u,i}$ to indicate the overall satisfaction of a user $u$ for the item $i$. Besides, each item is accompanied with a set of images $\mathcal{P}_i$, which visually depicts different aspects of the item,  e.g., \emph{``food"} and \emph{``inside ambience"} for restaurants. The primary goal is to predict the unknown ratings of items that the users have not reviewed yet. A review $d_{u,i}$ is a piece of text which describes opinions of a user $u$ on different aspects $a \in \mathcal{A}$ towards an item $i$. An aspect here is an attribute of items that has commented on in a review, such as \emph{``food"} for \emph{restaurants}. In this paper, we only consider the case that all the items are from the same category, i.e., they share the same set of aspects $\mathcal{A}$. Aspects that users care for items are latent and learned from reviews by our proposed topic model, in which each aspect is represented as a distribution of the same set (e.g., $K$) of latent topics. Table~\ref{tab:notation} lists the key notations. Before introducing our method, we would like to first clarify the  concepts of \emph{aspects}, \emph{latent topics}, and \emph{latent factors}.

\begin{itemize}[align=left,style=nextline,leftmargin=*,labelsep=\parindent,font=\normalfont]
	\item \textbf{Aspect} - it is a high-level semantic concept, which represents the attribute of items that users commented on in reviews,  such as \emph{``food''} for \emph{restaurant} and \emph{``battery"} for \emph{mobile phones}.
	
	\item \textbf{Latent topic  \& latent factor} - in our context, both concepts represent a more fine-grained concept than \emph{``aspect"}. A latent topic or factor can be regarded as a \emph{sub-aspect} of an item. For instance, for the ``food" aspect, a related latent topic could be ``\emph{breakfast}" or ``\emph{Italian cuisine}". We adopt the terminology of \emph{latent topic} in topic models and \emph{latent factor} in matrix factorization.  Accordingly, ``latent topics" are discovered by topic model on reviews, and ``latent factors" are learned by matrix factorization on ratings.
\end{itemize}

\subsection{Aspect-aware Latent Factor Model}
Based on the observations that (1) different users may care about different aspects of an item and (2) users' preferences may differ from each other for the same aspect, we claim that the overall satisfaction of a user $u$ towards an item $i$ (i.e., the overall rating $r_{u,i}$) depends on $u$'s satisfaction on each aspect $a$ of $i$ (i.e., \emph{aspect rating} $r_{u,i,a}$) and the importance of each aspect (of $i$) to $u$ (i.e., \emph{aspect importance} $\rho_{u,i,a}$). Based on the assumptions, the overall rating $r_{u,i}$ can be predicted as:
\begin{equation}
    \hat{r}_{u,i} = \sum_{a\in\mathcal{A}}  \overbrace{\rho_{u,i,a}}^{\xsub{aspect importance}} \underbrace{r_{u,i,a}}_{\xsub{aspect  rating}}
\end{equation}

\subsubsection{Aspect rating estimation.}
Aspect rating (i.e., $r_{u,i,a}$) reflects the satisfaction of a user $u$ towards an item $i$ on the aspect $a$. To receive a high aspect rating $r_{u,i,a}$, an item should at least possess the characteristics/attributes that the user is interested in on this aspect. Moreover,  the item should satisfy user's expectations on these attributes in this aspect. In other words, the item's attributes on this aspect should be of high quality such that the user likes it.  Take the ``food" aspect as an example, for a user who likes Chinese cuisine, to receive a high rating on the \emph{``food"} aspect from the user, a restaurant should provide Chinese dishes and the dishes should suit the user's tastes. Based on user reviews and item images, we could learn users' preferences and items' characteristics on each aspect and measure \emph{how the attributes of an item $i$ on aspect ``$a$" suit a user $u$'s requirements on this aspect}, denoted by $s_{u,i,a}$. We compute $s_{u,i,a}$ based on results of the proposed topic model MTAM (described in Sect.~\ref{sec:matm}), in which user's preferences and item's characteristics on each aspect are modeled as multinominal distributions of latent topics, denoted by $\bm{\theta_{u,a}}$ and $\bm{\psi_{i,a}}$, respectively. $s_{u,i,a}$ is then computed based on the Jensen–Shannon divergence~\cite{endres2003new} between $\bm{\theta_{u,a}}$ and $\bm{\psi_{i,a}}$:
\begin{equation} \label{eq:jsd}
s_{u,i,a} = 1-JSD(\bm{\theta_{u,a}}, \bm{\psi_{i,a}})
\end{equation}
Notice that a high value of $s_{u,i,a}$ does not mean a high rating $r_{u,i,a}$ - an item providing all the features that a user $u$ requires does not mean that it satisfies $u$'s expectations, since the provided ones could be of low quality. For instance, a restaurant provides all the Chinese dishes the user $u$ likes (i.e., high score $s_{u,i,a}$), but these dishes taste bad from $u$'s opinion (i.e., low rating $r_{u,i,a}$).

To accurately model the aspect rating, we refer to user's overall ratings on items.  Relying solely on user-item ratings, matrix factorization (MF)~\cite{koren2009matrix} could estimate the overall rating of users on unrated items. Here, we extend it to model aspect ratings. In MF,  users and items are mapped into a latent factor space, in which user's preferences and item's characteristics are modeled by  $f$ latent factors (i.e., $\bm{p_u} \in \mathbb{R}^{f \times 1}$ and $\bm{q_i} \in \mathbb{R}^{f \times 1}$).   The dot product of
the user's and item's vectors ($\bm{p_u}^T\bm{q_i}$) characterizes the user's overall interests on the item's characteristics, and is thus used to predict the rating of  $u$ to $i$. To enable MF to predict aspect ratings, we introduce a binary matrix  $\bm{W} \in \mathbb{R}^{f \times A}$ to associate the latent factors to different aspects, where $A$ is the number of aspects considered. We call this model  aspect-aware latent factor model (ALFM) model, in which the weight vector $\bm{w_a}$ in the $a$-th column of $\bm{W}$ indicates which factors are related to the aspect $a$. Thus, $\bm{p_{u,a}} = \bm{w_a} \odot \bm{p_u}$ denotes user's interests on aspect $a$ in the latent space, where $\odot$ represents element-wise product between vectors. Therefore, $(\bm{p_{u,a}})^T(\bm{q_{i,a}})$ represents the aspect rating of user $u$ to item $i$ on aspect $a$. To correlate the \emph{latent aspects }in ALFM to the \emph{explainable aspects} that users discuss in their reviews (e.g., food), we integrate the matching results of aspects based on user reviews (i.e., $s_{u,i,a}$) into ALFM:
 \begin{equation}
 \small
    r_{u,i,a} = s_{u,i,a} \cdot (\bm{w_a} \odot \bm{p_u})^T(\bm{w_a} \odot \bm{q_i})
 \end{equation}
 As a large $r_{u,i,a}$ requires  large values of both $s_{u,i,a}$ and $(\bm{w_a} \odot \bm{p_u})^T(\bm{w_a} \odot \bm{q_i})$, we expect that the results learned from reviews could guide the learning of latent factors.

\subsubsection{Aspect importance estimation.}
We rely on user reviews to estimate $\rho_{u,i,a}$, as users often discuss their interest topics of aspects in reviews, such as different \emph{cuisines} in the \emph{food} aspect, in which a certain type of \emph{cuisine} can be regarded as a \emph{topic}. In general, the more a user comments on an aspect in reviews, the more important this aspect is to the user. Thus, we estimate the importance of an aspect according to the possibility of a user writing review comments from this aspect. When writing a review,  some users tend to write comments from the aspects according to their own preferences, while others like commenting on the most notable features that the item possesses. With these considerations, we use (1) $\pi_u$ to denote the probability of user $u$ commenting an item from his own preferences and (2) $\lambda_{u,a}$ ($\sum_{a\in\mathcal{A}}\lambda_{u,a}=1$) to denote the probability of user $u$ commenting on the aspect $``a"$ from his preferences. Accordingly, $(1-\pi_{u})$ denotes the probability of the user commenting from the item $i$'s characteristics ($\sum_{a\in\mathcal{A}}\lambda_{i,a}=1$), and $\lambda_{i,a}$ is the probability of user $u$ commenting item $i$ from the item's characteristics on the aspect $a$. Thus, the probability of a user $u$ commenting an item $i$ on an aspect $a$ (i.e., $\rho_{u,i,a}$) is:
\begin{equation} \label{eq:rho}
	\rho_{u,i,a} = \pi_{u}\lambda_{u,a} + (1-\pi_{u})\lambda_{i,a}
\end{equation}
$\lambda_{u,a}$, $\lambda_{i,a}$, and  $\pi_u$ are estimated by the proposed generative probabilistic model - MTAM, which simulates the generation process of a user writing a review, as detailed in Sect.~\ref{sec:matm}.


\subsection{Multi-modal Aspect-aware Topic Model} \label{sec:matm}
\begin{figure}
    \centering
        \includegraphics[width = 10cm]{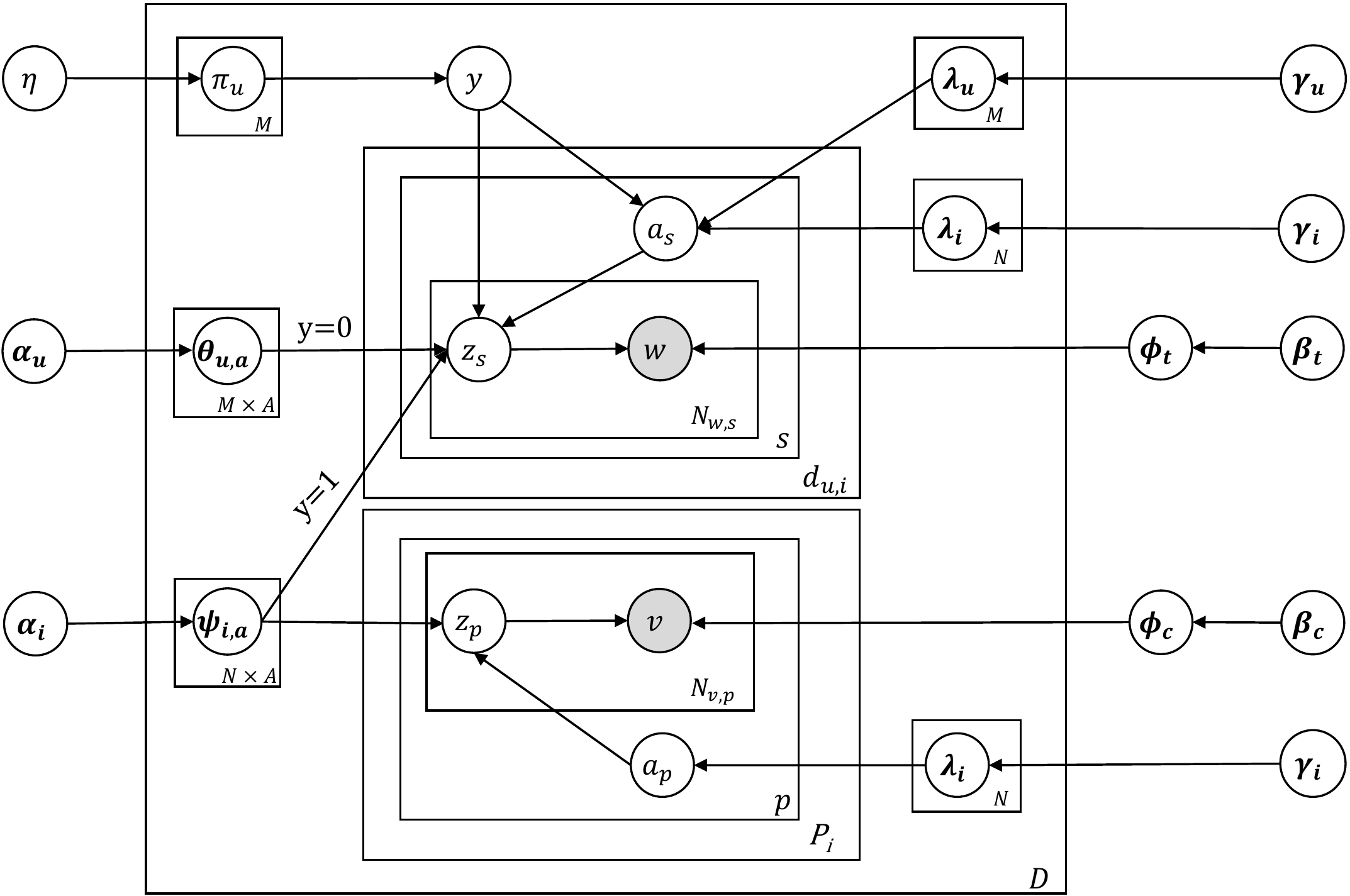}
   \caption{The graphical representation of the MMATM model. }
       \label{fig:matm}
\end{figure}
We assume that  a set of latent topics (i.e., $K$ topics) covers all the topics that users discuss in the reviews. The interests of a user $u$ on a specific aspect $a$ is represented by $\bm{\theta_{u,a}}$, which is a multinomial distribution of the latent topics. Similarly, the characteristics of an item $i$ on a specific aspect $a$ is represented by $\bm{\psi_{i,a}}$, which is also a multinomial distribution of the same set of latent topics.
The corpus $\mathcal{D}$ contains reviews of users towards items $d_{u,i}$ and images of items $\mathcal{P}_i$ for all $u \in \mathcal{U}$ and $i \in \mathcal{I}$. $\bm{\theta_{u,a}}$ is determined  based on text reviews, and $\bm{\psi_{i,a}}$ is affected by both text reviews and $i$'s visual content in $\mathcal{P}_i$. A latent topic is a multinomial distribution of text words (in reviews) and visual words (in items' images). Detailed information about visual words and our implementation of visual words are specified in Section 4.1.1. Based on these assumptions, we propose a multi-modal aspect-aware topic model (MATM for short) to model $\bm{\theta_{u,a}}$, $\bm{\psi_{i,a}}$, $\bm\lambda_{u}$, $\bm\lambda_{i}$, and $\pi_u$ by simulating the generation of the corpus $\mathcal{D}$.

The graphical representation of MATM is shown in Fig.~\ref{fig:matm}. In the figure, the shaded circles indicate observed variables, while the unshaded ones represent the latent variables. $a_s$ and $a_p$ denote the aspects assigned to a sentence $s$ and an image $p$, respectively. Notice that a sentence
usually discusses the same topic $z$, which could be from user’s preferences or from item’s characteristics. Therefore, to mimic the generation process of writing a review, an indicator variable  $y \in \{0,1\}$  is introduced to decide the topic $z_s$ for a sentence $s$. The indicator variable $y$ is  parameterized by $\pi_u$ based on a Bernoulli distribution. Specifically,  when $y=0$, the sentence is generated based on user's preference $\bm{\theta_{u,a}}$; otherwise, it is generated according to $\bm{\psi_{i,a}}$.  $\pi_u$ is user-dependent, as the tendency to comment from $u$'s personal preferences or from the item's characteristics is determined by $u$'s personality. As item images depict the characteristics of items, the generation of images (of an item) are decided based on the aspect distribution of this particular item. Hence, the topic $z_p$ for an image $p$ is generated according to $\bm{\psi_{i,a}}$.\footnote{Here we assume that an image depicts the characteristics of an item on the same topic $z_p$, which is similar to the assumption that a sentence focuses on the same topic.}

The generation process of MATM is shown in  Algorithm~\ref{alg:geneproc}. Let $a_s$ denote the aspect assigned to a sentence $s$.  If $y=0$, $a_s$ is drawn from $\bm{\lambda_u}$ and $z_s$ is then generated from $u$'s preferences on aspect $a_s$: $\bm{\theta_{u,a_s}}$; otherwise, if $y=1$,  $a_s$ is drawn from $\bm{\lambda_i}$ and $z_s$ is then generated from $i$'s characteristics on aspect $a_s$: $\bm{\psi_{i,a_s}}$. Then all the words $w$ in sentence $s$ is generated from $z_s$ according to the word distribution: $\bm{\phi_{z_s, w}}$. Similarly, let $a_p$ denote the aspect assigned to an image $p$ (of the item $i$). $a_p$ is drawn from $\bm{\lambda_i}$ and $z_p$ is then generated from $i$'s characteristics on aspect $a_p$: $\bm{\psi_{i,a_p}}$. Then all visual words $v$ of the image $p$ is generated from $z_p$ according to the visual word distribution: $\bm{\phi_{z_p, v}}$.
From the generation process, we can see that $\bm{\psi_{i,a}}$ is directly affected by text reviews of item $i$ and its visual content $v \in \mathcal{P}_i$. Notice that the visual content in images also affects the generation of text review, since it affects the distribution of $\bm{\psi_{i,a}}$, which generates text word $w$ when $y=1$. It makes sense since the images reflects items' characteristics from the visual appearances, which should affect users' preferences on items.

\begin{algorithm}
    \For{Each topic $k = 1, ..., K$ }
    {
        Draw $\bm{\phi_{k,t}} \sim Dir(\cdot|\bm{\beta_t})$\;
        Draw $\bm{\phi_{k,c}} \sim Dir(\cdot|\bm{\beta_c})$\;
    }
    \For{Each user $u \in \mathcal{U}$, each aspect $a \in \mathcal{A}$}
    {
        Draw $\bm{\theta_{u,a}} \sim Dir(\cdot|\bm{\alpha_u})$\;
    }
    \For{Each item $i \in \mathcal{I}$, each aspect $a \in \mathcal{A}$}
    {
        Draw $\bm{\psi_{i,a}} \sim Dir(\cdot|\bm{\alpha_i})$\;
    }
    \For{Each review $d_{u,i}, u \in \mathcal{U}, i \in \mathcal{I}$ }
    {
            \For{Each sentence $s \in d_{u,i}$ }
            {
                Draw $y \sim Bernoulli(\cdot|\pi_u)$\;
                \If{$y_s == 0$}{Draw $a_s \sim Multi(\bm{\lambda_u})$ and then draw $z_s \sim Multi(\bm{\theta_{u,a_s}})$ \;}
                \If{$y_s == 1$}{Draw $a_s \sim Multi(\bm{\lambda_i})$ and then draw $z_s \sim Multi(\bm{\psi_{i,a_s}})$\;}

               \For{Each text word $w \in s$ }
               {
                Draw $w \sim Multi(\phi_{z_s,w})$
                }
            }
     }
    \For{Each image $p \in \mathcal{P}_i, i \in \mathcal{I}$}
            {
                Draw $a_p \sim  Multi(\bm{\lambda_i})$ and then draw $z_p \sim Multi(\bm{\psi_{i,a_p}})$\;
                  \For{Each visual word $v \in p$ }
               {
                Draw $v \sim Multi(\phi_{z_p,v})$
                }
           }

    \caption{Generation Process of MATM}
    \label{alg:geneproc}
\end{algorithm}

In MATM, $\bm{\alpha_u}$, $\bm{\alpha_i}$, $\bm{\beta_t}$, $\bm{\beta_c}$, $\bm{\gamma_u}$ and $\bm{\gamma_i}$ are pre-defined hyper-parameters and set to be symmetric\footnote{In fact, for each $\bm{\theta_{u,a}}, u \in \mathcal{U}, a \in \mathcal{A}$ or $\bm{\psi_{i,a}}, i \in \mathcal{I}, a \in \mathcal{A}$, there should be a Dirichlet prior (i.e., $\alpha_{u,a}$ or $\alpha_{i,a}$). In the presentation and implementation, we set them to be the same, namely, $\forall u \in \mathcal{U}, a \in \mathcal{A}, \bm{\alpha_{u,a}} = \bm{\alpha_u}$ and $\forall i \in \mathcal{I}, a \in \mathcal{A}, \bm{\alpha_{i,a}} = \bm{\alpha_i}$.}. Parameters need to be estimated including $\bm{\theta_{u,a}}$, $\bm{\psi_{i,a}}$, $\bm{\lambda_u}$, $\bm{\lambda_i}$, $\pi_u$, $\bm{\phi_t}$, and $\bm{\phi_c}$. Different approximate inference methods have been developed for parameter estimation in topic models, such as variation inference~\cite{blei2003latent} and collapsed Gibbs sampling~\cite{griffiths2004}. We apply collapsed Gibbs sampling to infer the parameters, since it has been successfully applied in many large scale applications of topic models~\cite{cheng2016effective,qian2016multi}. In the Gibbs sampling process, the key step is to decide $z_s$ and $z_p$ for a each sentence $s$ and each image $p$ conditioned on all other variables.  Notice that $y_s$, and $a_s$  and $z_s$ must be sampled jointly, because $y_s$ decides to whether sample $a_s$ from $\bm{\lambda_u}$ or from $\bm{\lambda_i}$, and then subsequently decide the sampling of $z_s$ from $\bm{\theta_{u,a_s}}$ or $\bm{\psi_{i, a_s}}$.  Formally, we define that $\bm{S}$ is a sequence of sentences during the sampling process. $\bm{A}$, $\bm{Z}$ and $\bm{Y}$ denote the set of aspects $a$, topics $z$ and
indicators $y$ to the sentence sequence, respectively. $\bm{S_{ \neg s}}$ denotes $\bm{S}$ excluding
the sentence $s$. Similar notations are used for other variables. For $\bm{S} = \{s, \bm{S}_{\neg s}\}$,
$\bm{A} = \{a_{s}, \bm{A}_{\neg {s}}\}$, $\bm{Z} = \{z_{s}, \bm{Z}_{\neg {s}}\}$, and $\bm{Y} = \{y_s, \bm{Y}_{\neg s}\}$, the joint probability of
sampling $z_{s} = k$,  $y_s = 0$ and $a_s = a$ is:

\begin{equation}
\begin{split}
  &p(z_{s}=k, y_s=0, a_s=a| \bm{Z}_{\neg {s}}, \bm{A}_{\neg {s}}, \bm{Y}_{\neg s}, s, \bm{S}_{\neg s}, \bm{\alpha_u}, \bm{\alpha_i}, \bm{\beta_t},  \bm{\gamma_u},\bm{\gamma_i} , \bm\eta)\\
  &~~~~~~~~~~~~~~~~~~~~~~\propto (\eta_0 + N_{u,y_0,\neg s}) \cdot \frac{\gamma_{u,a} + N_{u, \neg s}^{a}}{\sum_{a=1}^A(\gamma_{u,a} + N_{u, \neg s}^a)} \cdot \frac{\alpha_{u,a,k} + N_{u,a, \neg s}^k}{\sum_{k=1}^K(\alpha_{u,a,k} + N_{u,a, \neg s}^k)} \cdot \frac{\beta_t + N_{k, \neg \bm{w_s}}^t}{\sum_{t=1}^T(\beta_t + N_{k, \neg \bm{w_s}}^t)} \label{eq:y0}
\end{split}
\end{equation}

Similarly, the joint probability of sampling $z_{s} = k$,  $y_s = 1$ and $a_s = a$  is:
\begin{equation}
\begin{split}
  &p(z_{s}=k, y_s=1, a_s=a| \bm{Z}_{\neg {s}}, \bm{A}_{\neg {s}}, \bm{Y}_{\neg s}, s, \bm{S}_{\neg s},\bm{\alpha_u},\bm{\alpha_i}, \bm{\beta_t}, \bm{\gamma_u},\bm{\gamma_i} , \bm\eta)\\
  &~~~~~~~~~~~~~~~~~~~~~~\propto (\eta_1 + N_{u,y_1,\neg s}) \cdot \frac{\gamma_{i,a} + N_{i, \neg s}^{a}}{\sum_{a=1}^A(\gamma_{i,a} + N_{i, \neg s}^a)} \cdot \frac{\alpha_{i,a,k} + N_{i,a, \neg s}^k}{\sum_{k=1}^K(\alpha_{i,a,k} + N_{i,a, \neg s}^k)} \cdot \frac{\beta_t + N_{k, \neg \bm{w_s}}^t}{\sum_{t=1}^T(\beta_t + N_{k, \neg \bm{w_s}}^t)} \label{eq:y1}
\end{split}
\end{equation}
 where $N_{u}^a$ denotes the number of times that the aspect $a$ is observed in user $u$'s reviews; and $N_{i}^a$ denotes the number of times that the aspect $a$ is observed in item $i$'s reviews. $``A"$ denotes the total number of aspects. $N_{u, \neg s}^a$ denotes the number of times aspect $a$ is observed in $u$'s review by excluding the sentence $s$. Similar definition is used for all notations in the form of $N_{\cdot, \neg s}^\cdot$.  $N_{u,a}^k$ is the number of observations of topic $k$ in the aspect $``a"$ of user $u$; and $N_{i,a}^k$ is the number of observations of topic $k$ in the aspect $``a"$ of item $i$. $N_{i,a, \neg s}^k$ denotes the number of observations of topic $k$ assigned to the aspect $a$ by excluding the sentence $s$.  $N_k^t$ denotes the number of times that term $t$ observed in topic $k$.  $N_{k, \neg \bm{w_s}}^t$ denotes the number of times that term $t$ observed in topic $k$ by excluding words $\bm{w_s}$ in the sentence $s$. $\bm{w_s}$ denotes all the words in sentence $s$. Term $t$ refers to the element of a vocabulary. $T$ is the vocabulary size of text words. $N_{u,y_0}$ and $N_{u,y_1}$ denote that for a particular user $u$: the number of times that words (in $u$'s reviews) are drawn from his/her preferences or from item's characteristics, respectively.

The sampling of $z_p=k$ is independent of $y$. The probability of sampling $a_p=a$ and $z_p = k$ is:
\begin{equation}
\begin{split}
  &p(z_{p}=k, a_p = a|\bm{Z}_{\neg {v}}, \bm{A}_{\neg p}, p, \bm{P}_{\neg p}, \bm{\alpha_i},  \bm{\beta_c}, \bm{\gamma_i})~~~~~~~~~~~~~~~~~~~~ \\
  &~~~~~~~~~~~~~~~~~~~~~~~~~~~~~~~~~~~~~~~~~~~\propto \frac{\gamma_{i,a} + N_{i, \neg p}^{a}}{\sum_{a=1}^A(\gamma_{i,a} + N_{i, \neg p}^a)} \cdot \frac{\alpha_{i,a,k} + N_{i,a, \neg p}^k}{\sum_{k=1}^K(\alpha_{i,a,k} + N_{i,a, \neg p}^k)} \cdot \frac{\beta_c + N_{k, \neg \bm{v_p}}^c}{\sum_{c=1}^C(\beta_c + N_{k, \neg \bm{v_p}}^c)}\label{eq:y1}
 \end{split}
\end{equation}
Similar to the definition of term $t$, $c$ is a distinct visual word in the vocabulary of visual words. $C$ is the size of visual word vocabulary. $a_{p}$ denotes the assigned aspects for image $p$. $\bm{v_p}$ denotes all the visual words in the image $p$. The definition of $N_{k, \neg \bm{v_p}} ^c$ is the number of times of visual term $c$ assigned to the topic $k$ after excluding all the visual words $\bm{v_p}$ in the image $p$. $N_k^c$ is the number of terms that visual term $c$ is drawn from topic $k$. Based on the state of the Markov chain $\bm{Y}$ and $\bm{Z}$, we can estimate the parameters:
\begin{align*}
 &\theta_{u,a,k} = \frac{\alpha_{u,a,k} + N_{u,a}^k}{\sum_{k=1}^K(\alpha_{u,a,k} + N_{u,a}^k)};
 \quad
 \lambda_{u,a} = \frac{\gamma_{u,a} + N_{u}^a}{\sum_{a=1}^A(\gamma_{u,a} + N_{u}^a)};
 \quad
 \phi_{k,t} = \frac{\beta_t + N_k^t}{\sum_{t=1}^T(\beta_t + N_k^t)};
 \\
 & \psi_{i,a,k} = \frac{\alpha_{i,a,k} + N_{i,a}^k}{\sum_{k=1}^K(\alpha_{i,a,k} + N_{i,a}^k)};
  \quad
 \lambda_{i,a} = \frac{\gamma_{i,a} + N_{i}^a}{\sum_{a=1}^A(\gamma_{i,a} + N_{i}^a)};
 \quad
 \phi_{k,c} = \frac{\beta_c + N_k^c}{\sum_{c=1}^C(\beta_c + N_k^c)}\\
 &\pi_u = \frac{\eta_0 + N_{u,y_0}}{\eta_1 + \eta_0 + N_{u,y_1} + N_{u,y_0}}
\end{align*}
Notice that $N_i^a$ counts all the images $p \in \mathcal{P}_i$ and the review sentences $s$ of item $i$ which are assigned aspects from the characteristics of item $i$, which indicates that $\lambda_{i,a}$ is decided by both the item $i$'s textual reviews and its images.  Besides,  $N_{i,a}^k = \sum_{t=1}^T N_{i,a,t}^k + \sum_{c=1}^C N_{i,a,c}^k$, namely, $N_{i,a}^k$ is the total number of times that both the textual words and visual words of item $i$ are assigned to topic $k$ in the aspect $``a"$. Therefore, $\bm{\psi_{i,a}}$ is affected by both the textual reviews and related images.

\subsection{Model Inference} With the results of MATM, $\rho_{u,i,a}$ and $s_{u,i,a}$ can be computed using Eq.~\ref{eq:rho} and ~\ref{eq:jsd}, respectively. With the consideration of bias terms (i.e., $b_u, b_i, b_0$) in ALFM, the overall rating can be estimated as\footnote{In our experiments, we tried to normalize $\rho_{u,i,a}$ or $\rho_{u,i,a} \cdot s_{u,i,a}$ in Eq.~\ref{eq:re2}, while no improvement has been observed.},
\begin{equation} \label{eq:re2}
	\hat{r}_{u,i} = \sum_{a\in\mathcal{A}}(\rho_{u,i,a}\cdot s_{u,i,a} \cdot (\bm{w_a} \odot \bm{p_u})^T(\bm{w_a} \odot \bm{q_i}) )  + b_u + b_i + b_0
\end{equation}
where $b_0$ is the average rating,  $b_u$ and $b_i$ are the user and item biases, respectively. The learning of the parameters is to minimize the rating prediction error in the training dataset. The optimization objective function is
\begin{equation} \label{eq:ojf}
    	\underset{p*,q*}{\text{min}} \frac{1}{2}\sum_{u,i} (r_{u,i}  -\hat{r}_{u,i})^2 + \frac{\mu_u}{2} ||\bm{p_u}||_2^2 + \frac{\mu_i}{2} ||\bm{q_i}||_2^2  + \mu_w \sum_a||\bm{w_a}||_1 + \frac{\mu_b}{2} (||b_u||_2^2 + ||b_i||_2^2)
\end{equation}
where $||\cdot||_2$ denotes the $\ell_2$ norm for preventing model overfitting, and $||\cdot||_1$ denotes the $\ell_1$ norm.  $\mu_u, \mu_i, \mu_w$, and $\mu_b$ are regularization parameters, which are tunable hyper-parameters. In practice, we relax the binary requirement of $\bm{w_a}$ by using $\ell_l$ norm. It is well known that  $\ell_l$ regularization yields sparse solution of the weight~\cite{mairal2010online}. The $\ell_2$ regularization of $\bm{p_u}$ and $\bm{q_i}$ prevents them to have arbitrarily large values, which would lead to arbitrarily small values of $\bm{w_a}$.

\textbf{Optimization.} We use the stochastic gradient descent (SGD) to learn the parameters by optimizing the objective function Eq.~\ref{eq:ojf}. In each step of SGD, the localized optimization is performed on a rating $r_{u,i}$.  Let $L$ denote the loss, and the gradients of parameters are given as follows:
    \begin{align*}\label{eq:user}
		&\frac{\partial L}{\partial p_u}=\sum_{i=1}^{N}(\sum_{a}\rho_{u,i,a}s_{u,i,a}w_a^2)(\hat{r}_{u,i}-r_{u,i})q_i + \mu_u p_u \\
        &\frac{\partial L}{\partial q_i}=\sum_{u=1}^{M}(\sum_{a}\rho_{u,i,a}s_{u,i,a}w_a^2)(\hat{r}_{u,i}-r_{u,i})p_u + \mu_i q_i \\
        &\frac{\partial L}{\partial w_a}=\sum_{u=1}^{M}\sum_{i=1}^{N}\rho_{u,i,a}s_{u,i,a}(\hat{r}_{u,i}-r_{u,i})p_uq_iw_a + \frac{\mu_w w_a} {\sqrt{(w_a^2+\epsilon)}}
    \end{align*}
Here, we omit the gradients of $b_u$ and $b_i$, as they are the exactly same as in the standard biased MF~\cite{koren2009matrix}.  $M$ and $N$ are the total number of users and items in the dataset.  Notice that in the deriving of the gradient for $w_a$, we use $\sqrt{w_a^2+\epsilon}$ in place of $||w_a||_1$, because $\ell_1$ norm is not differentiable at 0. $\epsilon$ can be regarded as a ``smoothing parameter" and is set to $10^{-6}$ in our implementation.  We use Bold Driver~\cite{gemulla2011large} to  adjust the learning rate in each iteration: increasing the learning rate by $5\%$ if error rate is reduced; otherwise, resetting it to the value of the previous iteration and decreasing it by $50\%$.

\subsection{Discussion}  In our model, the users' interests and items' characteristics learned from user reviews and item images are integrated into ALFM to guide the learning of aspect-aware latent factors, such that ALFM could estimate aspect ratings for the aspects discussed in reviews, which makes the predicted rating explainable. Besides, by modeling aspect rating and aspect importance simultaneously, our model can achieve better performance on rating prediction.  Furthermore, our model correlates the latent topics learned from reviews and latent factors learned from ratings on the ``aspect" level, which is very different from previous review-based rating prediction methods (e.g., HFT~\cite{mcauley2013hidden}, CTR~\cite{wang2011collaborative}, topicMF~\cite{bao2014topicmf}, RMR~\cite{ling2014ratings}, and ITLFM~\cite{zhang2016integrating}). In these models, latent topics are directly correlated to latent factors based on an \emph{one-to-one} correspondence relation, and thus the number of latent factors must be equal to the number of latent topics. On the contrary, our model has the flexibility of tuning the number of latent factors and the number topics separately on modeling ratings and reviews, respectively.

\begin{table}[]
		\centering
		\caption{Statistics of the evaluation datasets.}
		\begin{tabular}{|l|c|c|c|c|} \hline
Datasets	&	\# users	&	\# items	&	\# rates	&	Sparsity	\\ \hline
Beauty	&	22,363	&	12,101	&	198,502	&	0.9993	\\
CDs, \& Vinyl	&	75,258	&	64,421	&	1,097,597	&	0.9998	\\
Clothing, Shoes, \& Jewelry	&	39,387	&	23,033	&	278,677	&	0.9997	\\
Movies \& TV	&	123,960	&	50,052	&	1,697,533	&	0.9997	\\
Cell Phones \& Accessories	&	27,879	&	10,429	&	19,439	&	0.9999	\\
Yelp	&	7,109	&	1,416	&	135,015	&	0.9866	\\ \hline

		\end{tabular}
		\label{tab:dataset}
\end{table}

\section{Experimental Setup}\label{sec:expconfig}
To evaluate the effectiveness of the proposed model, we conducted comprehensive experimental studies on real-word datasets collected from Amazon and Yelp. The datasets used in our experiments are publicly accessible, which will be detailed in the next subsection. As our model is delicately designed for rating prediction, the rating prediction will be the main evaluation task in our experiments. Besides, we will also evaluate its performance on top-n recommendation. In summary, our experiments mainly answer the following questions:
\begin{itemize} [align=left,style=nextline,leftmargin=*,labelsep=\parindent,font=\normalfont]
\item \textbf{RQ1:} How do different numbers of  latent factors and latent topics affect the performance of our model? More importantly, is the setting of $f=K$ optimal, which is a default assumption for many previous models? (Sect.~\ref{sec:modelanalysis})
\item \textbf{RQ2:} Is image content useful for capturing user's preferences? Could our model effectively integrate the visual information to model user's preferences?\footnote{Notice here our aim is to investigate whether the visual information can be integrated with reviews and ratings to improve the recommendation performance. We do not claim that the proposed model is optimal on leveraging image and reviews to model user preferences.} (Sect.~\ref{sec:effectip})
\item \textbf{RQ3:} Can our  model outperform the state-of-the-art rating prediction methods, which also consider both ratings and reviews? (Sect.~\ref{sec:comp})
\item \textbf{RQ4:} Could our model alleviate the cold-start problem when users have only few ratings? (Sect.~\ref{sec:coldstart})
\item \textbf{RQ5:} Can our model explicitly provide an interpretation of a high or low predicted rating? (Sect.~\ref{sec:interpret})
\item \textbf{RQ6:} Comparing to the state-of-the-art top-n recommendation algorithms, how does our model perform on the top-n recommendation task? (Sect.~\ref{sec:topn})
\end{itemize}
\subsection{Dataset} \label{sec:dataset}
We adopt the publicly accessible Amazon review dataset\footnote{http://jmcauley.ucsd.edu/data/amazon/} and Yelp Challenge dataset\footnote{http://www.yelp.com/dataset/challenge/} for experiments.

\begin{itemize} [align=left,style=nextline,leftmargin=*,labelsep=\parindent,font=\normalfont]
    \item \textbf{Amazon dataset.} This dataset was collected and released by~\cite{mcauley2013hidden}, which contains user interactions (review, rating, votes etc.) on items as well as the item metadata (e.g., description, price, brand, image URL, etc.) from Amazon. Each item is accompanied with an image. In this work, we leverage the review, rating, and image information. This dataset has been widely used for rating prediction with reviews and ratings in previous studies~\cite{mcauley2013hidden,ling2014ratings,tan2016rating,catherine2017transnets}. The dataset is organized into 24 product categories. In this paper, we used five categories (See Table~\ref{tab:dataset}) and focus on the 5-core version, with at least 5 reviews for each user or item.

    \item \textbf{Yelp dataset.} This dataset includes reviews of local business in 12 metropolitan areas across 4 countries. It provides a large number of user-item reviews and ratings; and more importantly, restaurant images. In this dataset, 64,000 images are labeled to four categories: \emph{food}, \emph{drink}, \emph{inside}, and  \emph{outside}, with 16,000 for each category.  We firstly select items with more than 10 images in those four categories, since our method considers items' visual features, and then further remove the items and users with less than 10 reviews. Therefore, this dataset is more dense compared to the  five Amazon datasets used in our experiments.
\end{itemize}

Some statistics of the datasets are shown in Table~\ref{tab:dataset}. For all the datasets, standard text pre-processing techniques are used to process the review, such as converting words to lowercase and stop-words removal. Besides, to filter noisy words in the reviews, terms appearing less than 10 times in the dataset are removed.

\subsubsection{Image Pre-processing}
In our topic model, each image is represented by a sequence of visual words (i.e., a visual-word-document). Specifically, a ``visual word” for an image is similar to a ``text word” in text document. In practice, an image is cut into blocks or patches and each block is regarded as a ``visual word”. For example, if we cut an image into $7\times7=49$ blocks, then this image can be represented by 49 ``visual words”. We concatenate the 49 visual words into a sequence; then the image is similar to a text document, which is a sequence of text words.

In the text domain, there is a text dictionary or vocabulary. Each text word can be indexed as a ``text term” in this vocabulary. Similarly, we need to generate a visual vocabulary for the “visual words” of images, so that each visual word can find a corresponding indexed “visual term” in this visual vocabulary. Usually, a clustering algorithm (e.g., K-means~\cite{macqueen1967some}) is used to generate the visual vocabulary. The procedure is as follows. Given a large set of images, (1) each image is cut into a certain number of blocks, e.g., 49 blocks, with the same size; (2) for each image block, we extract its visual features (e.g., ResNet features); (3) based on the visual features of image blocks, a clustering method (e.g., K-means clustering) is used to generate $K$ clusters. The centre of each cluster is regarded as a visual term (or each center is indexed into a visual term in the visual vocabulary).  To this end, to represent an image as a visual-word document, for each block of this image, we find its nearest centre and then replace this block by the corresponding visual term of this centre. In this way, an image can be represented by visual words.

In our implementation, each image is represented by $7 \times 7$ = 49 visual words.
Image features are extracted by the ResNet-152 (res5c)~\cite{he2016deep}.The outputs of the last convolutional layer in ResNet-152 (res5c) are used as the visual features for image patches. For each block, a 2048-dim visual feature vector is obtained. The K-means method is used for visual vocabulary generation and the vocabulary size is set to 4096 in our experiments.

\subsection{Comparative Methods} \label{sec:comp}
We compare the proposed model to the state-of-the-art rating prediction methods. It is worth noting that these methods are tuned on the validation dataset to obtain their optimal hyper-parameter settings for fair comparisons.

\begin{itemize} [align=left,style=nextline,leftmargin=*,labelsep=\parindent,font=\normalfont]
\item \textbf{BMF~\cite{koren2009matrix}.} It is a standard matrix factorization model with the consideration of biased terms. This method only leverage ratings only when modeling users' and items' latent factors.  It is typically a strong baseline model in collaborative filtering~\cite{koren2009matrix,ling2014ratings}. The regularization parameters are tuned to be 0.01.

\item \textbf{HFT~\cite{mcauley2013hidden}.} It is a pioneering model that combines reviews with ratings. HFT models ratings with a matrix factorization and the review text with latent topic model (e.g., LDA~\cite{blei2003latent}). We use it as a representation of methods, which use an exponential transformation function  to link the latent topics with latent factors, such as TopicMF~\cite{bao2014topicmf}.  The relative weight $\mu$ makes a trade-off between rating prediction error and likelihood of review text modeling, and is chosen to be 0.1 for better results.  The regularization parameters for this model are set to be 0.1 by tuning on validation  sets. The topic distribution can be modeled on either users or items. We use the topic distribution based on items, since it achieves better results. Note that in experiments, we add bias terms to HFT, which can achieve better performance.

\item \textbf{CTR~\cite{wang2011collaborative}.} This method also utilizes both review and rating information. It uses a topic model to learn the topic distribution of items ($\psi_i$), which is then used as the latent factors of items in MF with an addition of a latent variable $\epsilon_i$. We follow~\cite{wang2011collaborative} to tune the parameters.

\item \textbf{RMR~\cite{ling2014ratings}.} This method also uses both ratings and reviews. Different from HFT and CTR, which use MF to model rating, it uses a mixture of Gaussian to model the ratings.  We set hyperparameters $\alpha=0.1$, $\beta=0.01$, $\mu_0=0$, and $\sigma^2=1$ after trails on validation set.

\item \textbf{RBLT~\cite{tan2016rating}.} This method is a most recent method, which also uses matrix factorization to model ratings and LDA to model review texts. Instead of using an exponential transformation function to link the latent topics and latent factors (as in HFT~\cite{mcauley2013hidden} and TopicMF~\cite{bao2014topicmf}), this method linearly combines the latent factors (learned from ratings) and latent topics (learned from reviews) to represent users and items, with the assumption that the dimensions of topics and latent factors to be equal and in the same latent space. The same strategy is also adopted by ITLFM~\cite{zhang2016integrating}. Here, we use RBLT as a representative method for this strategy.

\item \textbf{EFM~\cite{zhang2014explicit}.} Different from above methods which rely on latent factor model to extract aspects of products, EFM applies external tools to extract specific product aspects and analyzes phrase-level sentiment on textual reviews to make recommendations. The extracted aspect-level features are then integrated with the collaborative filtering techniques to enhance the recommendation performance.

\item \textbf{TransNet~\cite{catherine2017transnets}.} This method adopts neural network frameworks for rating prediction. In this model, the reviews of users and items are used as input to learn the latent representations of users and items. The latent representations of a targeted user and a targeted item are concatenated and passed through a regression layer (consisting of Factorization Machine) to estimate the rating. This method is an early attempt to use reviews with ratings for rating prediction. We use the code published by the authors in experiments and tuned the parameters as described~\cite{catherine2017transnets} on validation set.

\item \textbf{TALFM.} It is a variant of our model. It only uses text reviews in MATM to model users' preference and items' characteristics. This model is also described in~\cite{cheng2017aspect}. The hyper-parameter settings are the same as MMALFM (see below).

\item \textbf{MMALFM} It is the proposed model, which uses both text reviews and item images in MATM. We set the hyperparameters $\alpha=\gamma = 0.1$, $\beta_w=\beta_v= 0.01$ in MATM, and the initial learning rate in ALFM to 0.01. For the regularization coefficients in ALFM: $\mu_u=\mu_i=\mu_b=0.1$ and $\mu_w=0.01$.  It takes around 50 iteration for ALFM to achieve convergence with Bold Driver~\cite{gemulla2011large}.
\end{itemize}

In our implementation, L-BFGS is used in the implementation of HFT, CTR, and RMR. We tune the number of latent factors ($f$) and the number of latent topics ($K$) in $[5, 10, 15, 20, 25]$. Note that $K$ and $f$ have to be the same value in HFT, CTR, RMR, and RBLT, and thus they are tuned together in these methods. For TALFM and MMALFM, $K$ and $f$ are tuned separately.

\subsection{Evaluation}
For each dataset, we randomly split it into training, validation, and testing set with ratio 80:10:10 for each user as in~\cite{mcauley2013hidden,ling2014ratings,catherine2017transnets}. Because we take the 5-core dataset where each user has at least 5 interactions, we have at least 3 interactions per user for training, and at least 1 interaction per user for validation and testing. Note that we only used the review information in the training set, because the reviews in the validation or testing set are unavailable during the prediction process in real scenarios. We use the standard root-mean-square error (\textbf{RMSE}) to evaluate to evaluate various models. Let $e=r-\hat{r}$ denote the prediction error,  RMSE is calculated as:
		\begin{equation}
	     RMSE = \sqrt{\frac{\sum_{i=1}^n e_i^2}{n}}
	  \end{equation}
where $n$ is the total number of predicted ratings in the testing set. A smaller value of RMSE indicates a better performance on the rating prediction task.

\section{Experimental Results} \label{sec:expresult}
This section reports the evaluation results. Firstly, we analyze the influence of the number of latent topics (i.e., \#topics) and the number of latent factors (i.e., \#factors) on the performance; then detailedly study the effects of item images in recommendation. In the next, we compare our model to the state-of-the-art methods on rating prediction. In particular, we evaluate the capability of our model on alleviating the problem of cold-start setting when only few ratings are available for users. In the next, we demonstrate the interpretability of our model. Finally, we report the performance of our model on the top-n recommendation task.

\subsection{Model Analysis}\label{sec:modelanalysis}
\subsubsection{Effects of \#factors \& \#topics (\textbf{RQ1})}
\begin{figure*}[]
		\centering
		\subfloat[Clothing]{
			\includegraphics[height = 5cm]{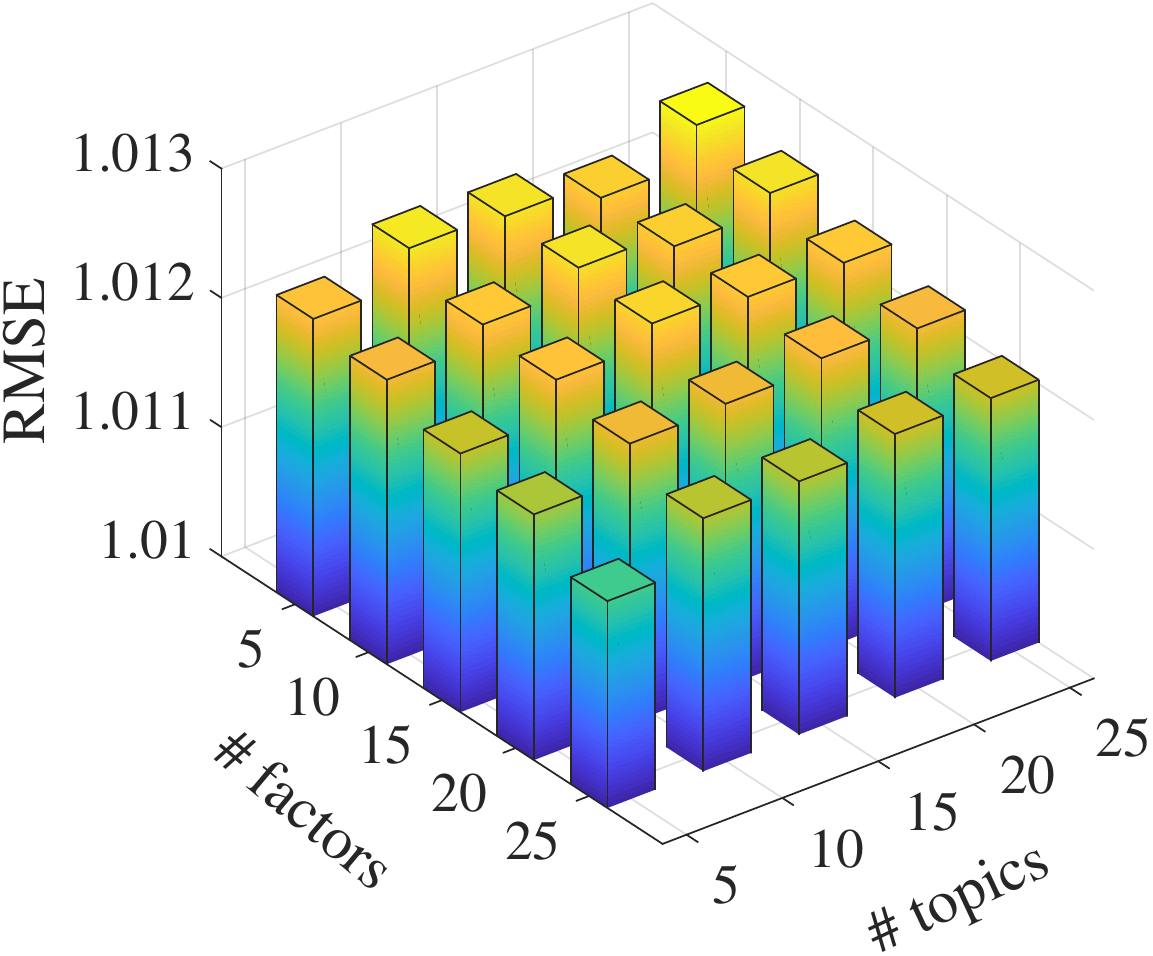}
            \label{fig:talfm}
		} \hspace{1cm}
		\subfloat[Yelp]{
			\includegraphics[height = 5cm]{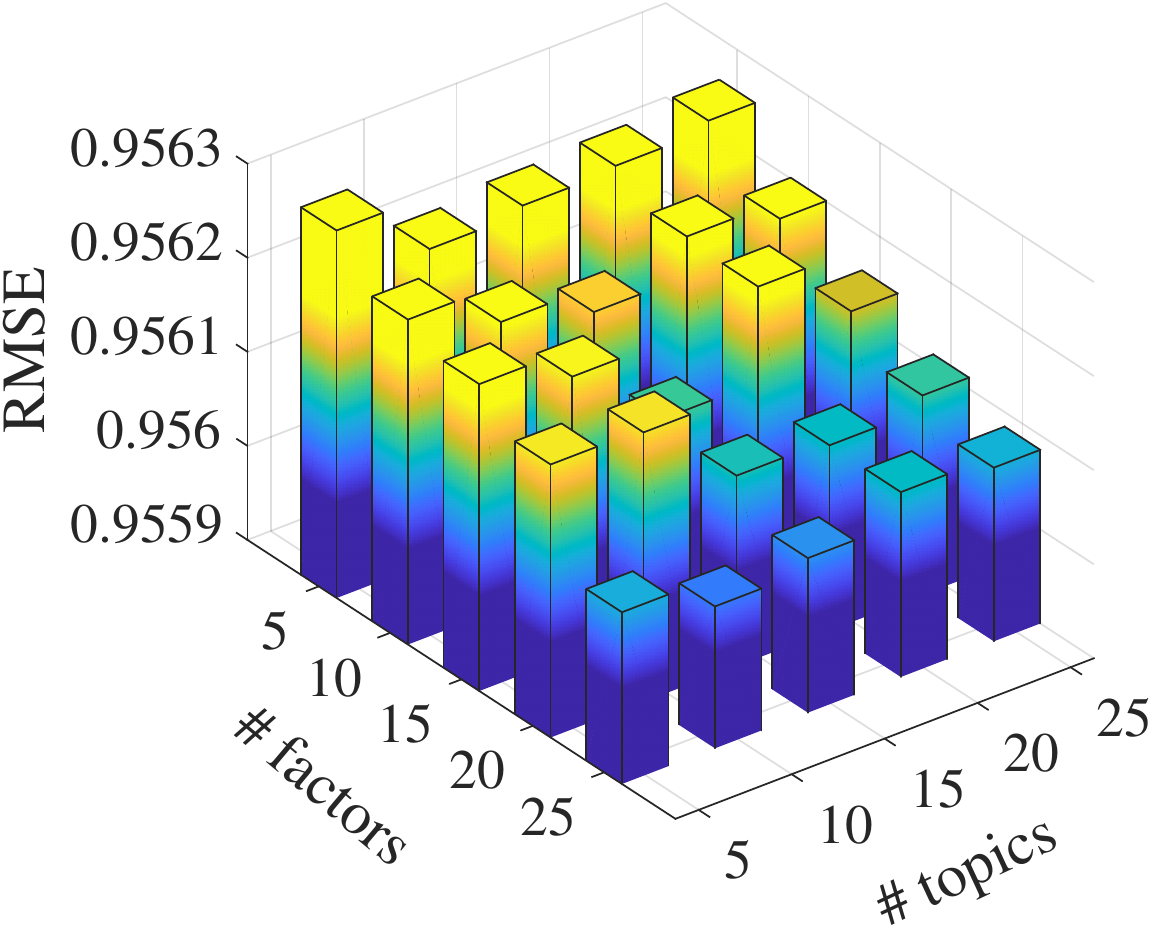}
            \label{fig:mmalfm}
		}

		\caption{Effects of \#factors v.s. \#topics.}
		\label{fig:factorvstopic}
	\end{figure*}

In matrix factorization, more  latent factors will lead to better performance unless overfitting occurs~\cite{he2016fast,koren2009matrix}; while the optimal number of latent topics in topic models (e.g., LDA) is dependent on the datasets~\cite{blei2012probabilistic,arun2010finding}. Accordingly, the optimal number of latent topics in topic model and the optimal number of latent factors in matrix factorization should be tuned separately. However, in the previous latent factor models (e.g., HFT, TopicMF~\cite{bao2014topicmf}, RMR, CTR, and RBLT) , the number of factors (i.e., \#factors) and the number of topics (i.e., \#topics) are assumed to be the same, and thus cannot be optimized separately. Since our model does not have such a constraint, we studied the effects of \#factors and \#topics individually. Fig.~\ref{fig:factorvstopic} shows the performance variations of MMALFM with the change of \#factors and \#topics. We use the RMSE on the Yelp dataset and Amazon Clothing dataset to illustrate the effects. Similar trends can be observed on other datasets.  From the figures, we can see that with the increase of \#factors, RMSE consistently decreases although the degree of decline is small. Notice that in our model, the rating prediction still relies on MF technique (Eq.~\ref{eq:re2}). Therefore, the increase of  \#factors could lead to better representation capability and thus more accurate prediction. In contrast, there is no general trends with the increase of \#topics, since the optimal number of topics is dependent on the training data. This also reveals that setting \#factors  and \#topics to be the same may not be optimal.

\begin{figure}
    \centering
        \includegraphics[width = 10cm]{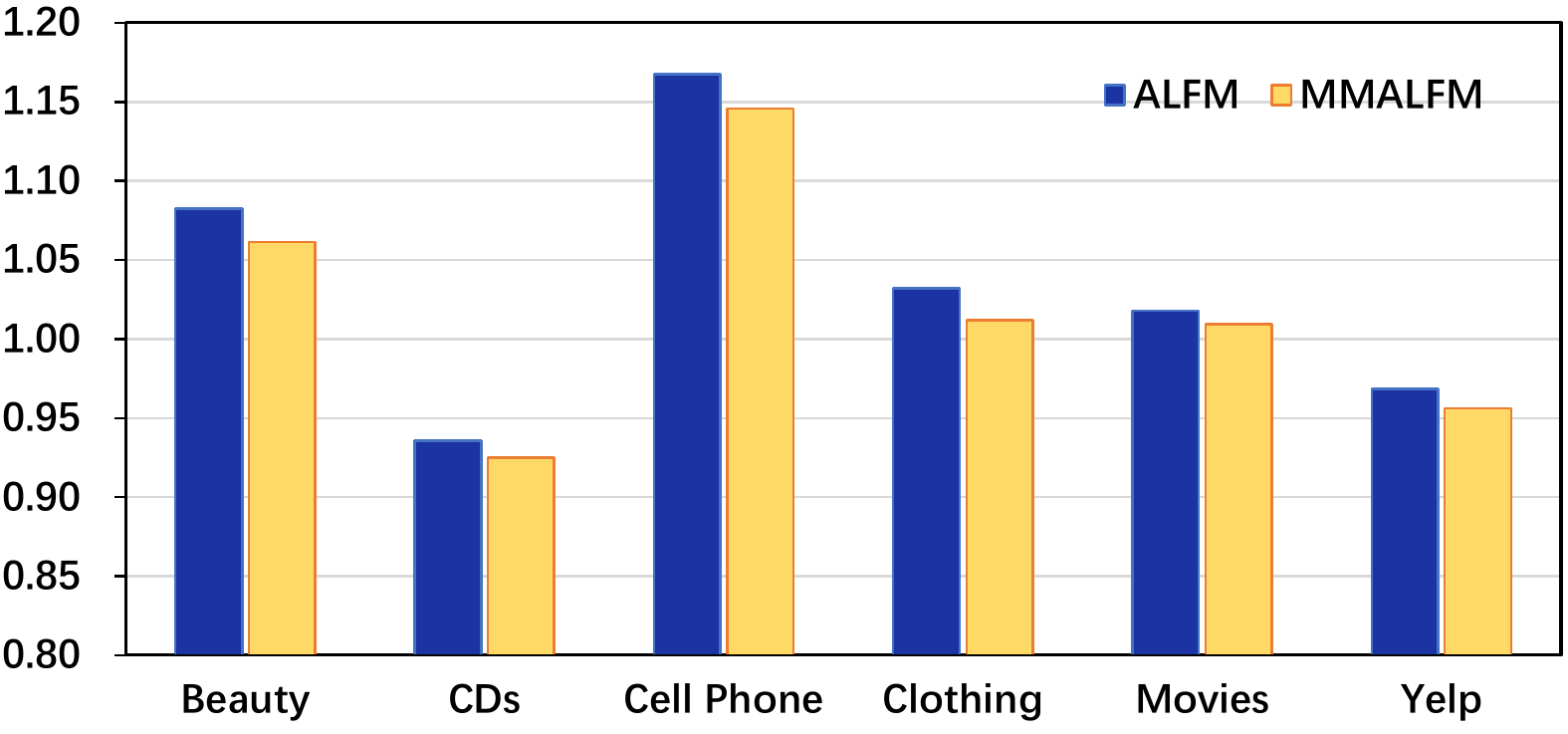}
   \caption{Effects of item images on the performance of RMSE based on $f=k=5$. }
       \label{fig:mmalfm-alfm}
\end{figure}
\subsubsection{Effects of Item Images (\textbf{RQ2})} \label{sec:effectip}
To demonstrate the effects of image on rating predication, we compare the results between TALFM and MMALFM, as shown in Figure~\ref{fig:mmalfm-alfm}. The concrete values of the performance can be found in the Table~\ref{tab:comp}.  From the results, we can see that with the consideration of image features, our model can consistently obtain better performance across different datasets. Because the textual reviews from users reveal semantic details about those products on different aspects, the improvements brought by images are not very large across those datasets. Generally, more important the visual appearance for the products, more improvements can be achieved with the additional consideration of images. For example, because users' preferences on clothes are largely affected by their visual features, the improvement on the ``Clothing" dataset is more obvious. By contrast, the improvements on some other domains (e.g., movies and CD) become limited, because the quality of products from those domains is hard to judge from images.

	\begin{figure}[]
			\centering
			\subfloat[Clothing]{
					\includegraphics[height = 4.5cm]{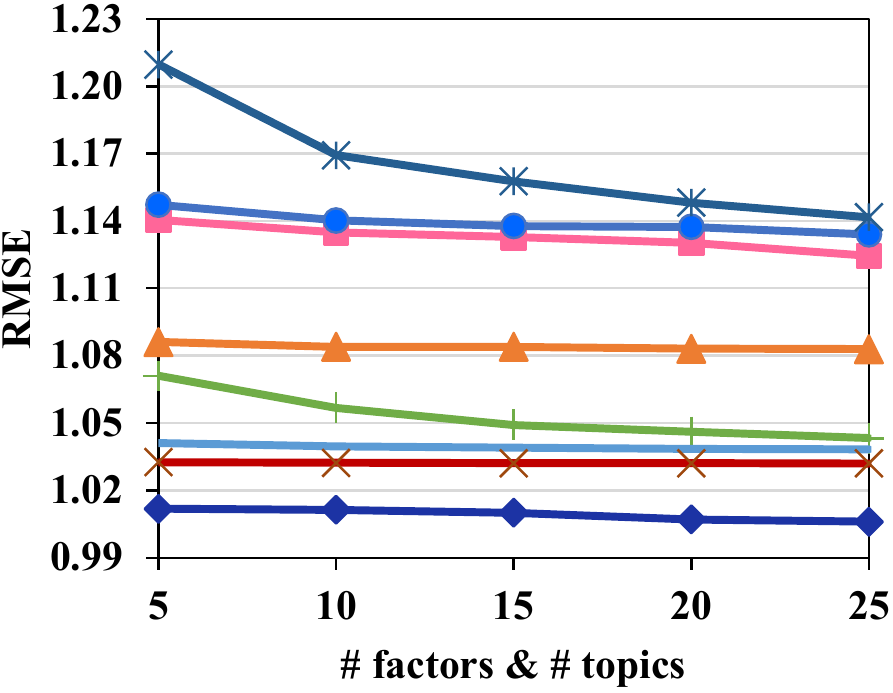}
			  	\label{fig:rmse}
			}
			\subfloat[Yelp]{
					\includegraphics[height = 4.5cm]{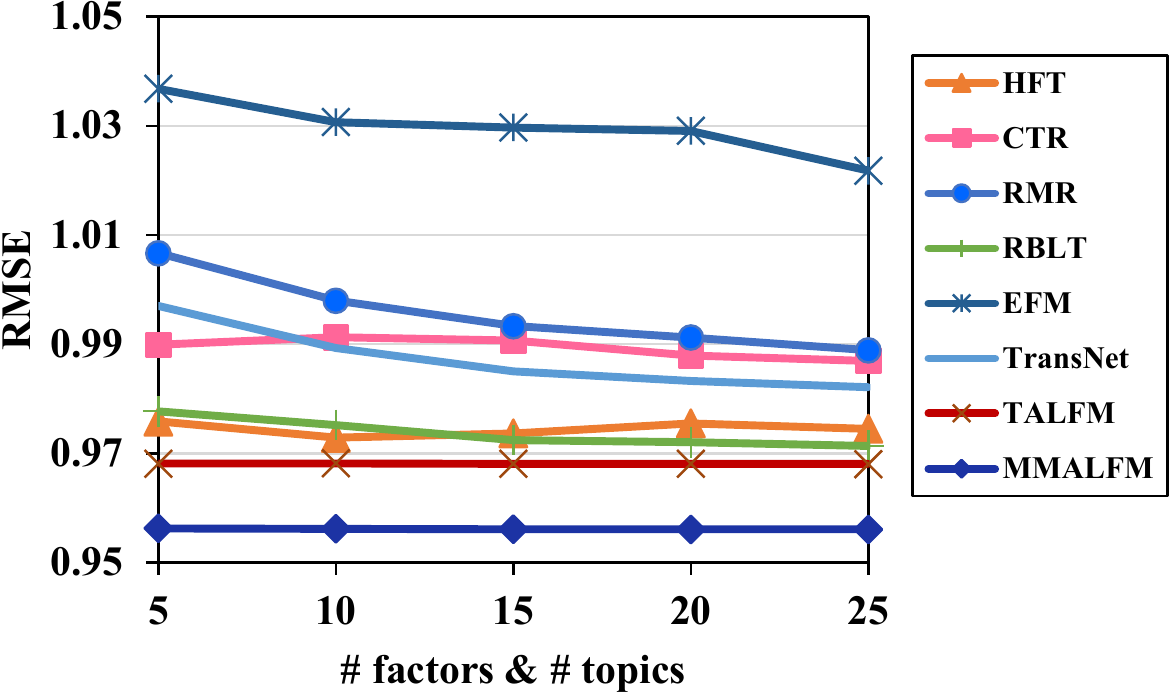}
					\label{fig:mae}
			}	
			\caption{Performance comparisons across different numbers of latent factors. }
			\label{fig:k}
	\end{figure}

\begin{table}[]
\small
 \centering
 \caption{Performance Comparisons in terms of RMSE  by $f=K=5$.}
     \begin{tabular}{|l|c|c|c|c|c|c|c|c|c|} \hline
Dataset	&	BMF	&	HFT	&	CTR	&	RMR	&	RBLT	&	EFM	&	TransNet	&	TALFM	&	MMALFM	\\ \hline \hline
Beauty	&	1.3395	&	1.1341	&	1.1704	&	1.1854	&	1.1133	&	1.2303 	&	1.2617	&	1.0825	&	\textbf{1.0613*} 	\\
CDs	&	1.1279	&	0.9446	&	0.9598	&	0.9811	&	0.9399	&	1.0173	&	0.9765	&	0.9359	&	\textbf{0.9250*} 	\\
Cell Phone	&	1.4326	&	1.2359	&	1.2694	&	1.2851	&	1.2207	&	1.3186	&	1.2863	&	1.1675	&	\textbf{1.1456*} 	\\
Clothing	&	1.2369	&	1.0829	&	1.1303	&	1.1473	&	1.0710	&	1.2098 	&	1.0410	&	1.0320	&	\textbf{1.0118*} 	\\
Movies	&	1.2249	&	1.0303	&	1.0362	&	1.0577	&	1.0188	&	1.1017 	&	1.0689	&	1.0180	&	\textbf{1.0091} 	\\
Yelp	&	1.1658	&	0.9759	&	0.9899	&	1.0067	&	0.9777	&	1.0368
 	&	0.9970	&	0.9682	& \textbf{0.9562*} 	\\ \hline	

    \end{tabular}
    \begin{tablenotes}
      \footnotesize
      \item ``*" denotes statistically significant differences ($p < 0.05$, a two-tailed paired t-test) with the performance of TALFM, which achieves the best performance in the remaining methods.
    \end{tablenotes}
    \label{tab:comp}
\end{table}

\subsection{Performance Comparison (\textbf{RQ3})}\label{sec:comp}
 Fig.~\ref{fig:k} shows the performance of the considered models on ``Clothing" and ``Yelp" datasets with the change of the number of latent factors (and topics). Similar results can be observed on other datasets.  Although $f$ (i.e., the number of latent factors) and $K$ (i.e., the number of latent topics) could be different in our models (i.e., TALFM and MMALFM),  $K$ and $f$ have to be the same value in HFT, CTR, RMR, and RBLT. For the ease of presentation, we present the results  of TALFM and MMALFM with settings of $K=f$ as other models in the figure.   Notice that in this figure, we have not presented the performance of BMF  for better visualization, because  its performance is much worse as shown in Table~\ref{tab:comp}. Table~\ref{tab:comp} shows the concrete scores obtained when $f=K=5$. From Fig.~\ref{fig:k}, we can observe the same trend: The performance of MMALFM is better than TALFM, due to the use of both text reviews and item images in preference modeling as discussed in the previous subsection. For the methods only considering ratings and text reviews, TALFM achieves the best performance, greatly outperforming all other methods.  In the remaining methods,  RBLT achieves the best performance, followed by the modified HFT, and then CTR, TransNet, RMR, and EFM. In general, with the increasing number of latent factors and latent topics, all models (except HFT) could achieve better performance.

Compared to BMF, which only uses ratings, our model achieves much better prediction performance. When $f=K=5$, the relative improvement of TALFM over BMF can achieve 17.52\% on average in terms of RMSE. Moreover, compared to the state-of-the-art methods using reviews, the relative improvements on average are 10.07\%, 6.68\%, 6.02\%, 5.11\%, 2.94\%,   and 2.04\% for EFM, RMR, TransNet, CTR, HFT, and RBLT, respectively. In those competitors, EFM relies on external tools to extract product aspects and analyze sentiments, and thus its performance will be affected by the performance of the used external tools. Although CTR, HFT, RMR and RBLT utilize topic models to automatically extract items' topic features from reviews as our model, they have not carefully modeled users' preferences on different aspects of items. The results demonstrate that our model is more effective in exploiting reviews and ratings, because it learns users' preferences and items' features in different aspects and is capable of estimating the aspect weights based on the targeted user’s preferences and targeted item’s features.

TransNet uses neural networks to learn user preferences and item features from reviews. Although neural networks have shown great capabilities in representation learning, the performance of TransNet is not very competitive in this case, which might be due to two reasons. Firstly, TransNet uses reviews as users' and items' representation input. However, there is a lot of noisy information in reviews, which would deteriorate the performance. Besides, when predicting unknown ratings, TransNet needs to generate a fake review, which is subsequently used to predict ratings. As a result, the error introduced by the generated fake review will also cause bias in the final performance.

\subsection{Cold-start Setting (\textbf{RQ4})} \label{sec:coldstart}
As shown in Table~\ref{tab:dataset}, the datasets are usually very sparse in practical systems. It is inherently difficult to provide satisfactory recommendation based on limited ratings. In the matrix factorization model, given a few ratings, the penalty function tends to push $\bm{p_u}$ and $\bm{q_i}$ towards zero. As a result, such users and items are modeled only with the bias terms~\cite{ling2014ratings}. Therefore, matrix factorization easily suffers from the cold-start problem. By integrating reviews in users' and items' latent factor learning, our model could alleviate the problem of cold-start to a great extent, since reviews contain rich information about user preferences and item features.

\begin{table}[]
\small
 \centering
 \caption{The percentage of cold-start users with different numbers (from 1 to 10) of training samples. The values in the first row indicate the specific number of training samples; the values in the second and third rows show the percentages of users with the corresponding training samples in the ``Clothing" and ``Yelp" datasets, respectively. }
     \begin{tabular}{|l|c|c|c|c|c|c|c|c|c|c|} \hline
\#training samples	&	1	&	2	&	3	&	4	&	5	&	6	&	7	&	8	&	9	&	10	\\ \hline \hline
Clothing (\%)	&	0.340 	&	2.399 	&	10.131 	&	22.095 	&	25.290 	&	13.472 	&	7.970 	&	4.729 	&	3.003 	&	2.292 	\\
Yelp (\%)	&	0.247 	&	1.575 	&	6.359 	&	15.069 	&	18.008 	&	11.856 	&	8.137 	&	5.719 	&	4.527 	&	3.544 	\\ \hline

    \end{tabular}
    \label{tab:coldstart}
\end{table}

	\begin{figure*}[]
			\centering
			\subfloat[Clothing]{
					\includegraphics[height = 4.5cm]{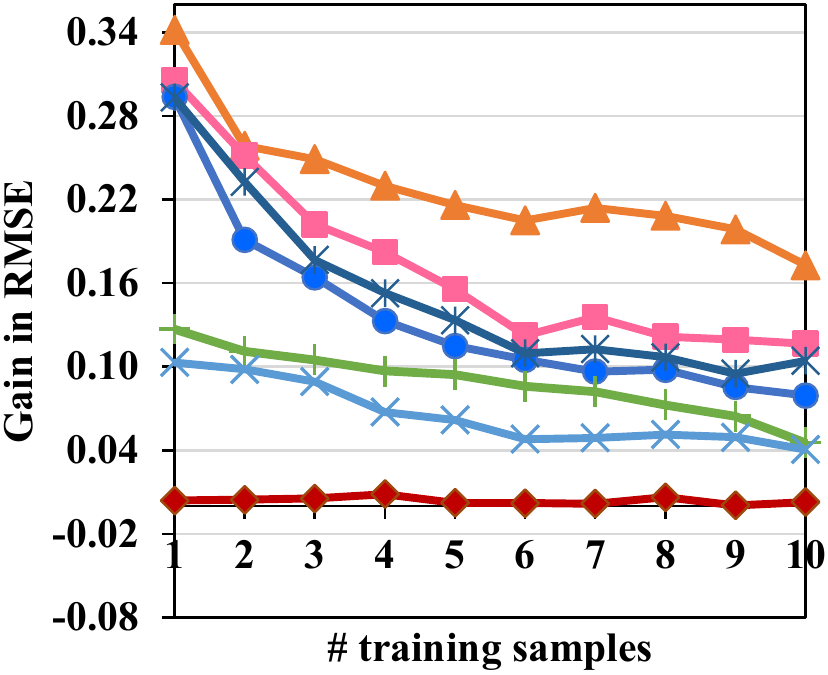}
            \label{fig:gain1}
			}
			\subfloat[Yelp]{
					\includegraphics[height = 4.5cm]{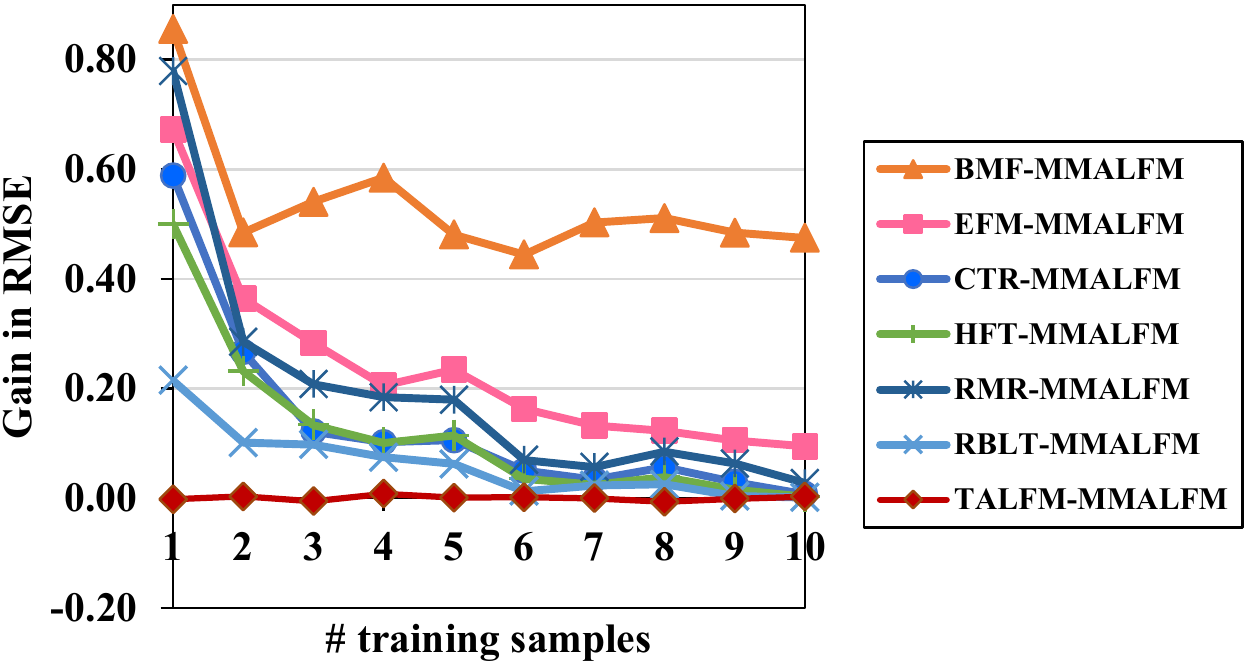}
             \label{fig:gain2}
			}		
            \caption{Gain in RMSE of MMTALM over baselines in the cold-start setting.}
			\label{fig:coldstart}

	\end{figure*}

To demonstrate the capability of our model on dealing with users with very limited ratings, we randomly split the datasets into training, validation, and testing sets in ratio 80:10:10 based on the number of ratings in each set. In this setting, it is not guaranteed  that a user has at least 3 ratings in the training set. It is possible that a user has no rating in the training set. For the users without any ratings in the training set, we also removed them in the testing set.  Then we evaluate the performance of users who have the number of ratings from 1 to 10 in the training set. We adopt the method in~\cite{tan2016rating} to demonstrate the capability of our model on cold-start setting.
In Fig.~\ref{fig:coldstart}, we show the \textbf{Gain in RMSE} ($y$-axis) grouped by the number of historical ratings ($x$-axis) of users in the testing set, which is equal to the average RMSE of baselines \emph{minus} that of our  model grouped by the number of ratings of users (e.g., ``BMF-MMALFM"). A positive value indicates that our model has better prediction. Fig.~\ref{fig:gain1} and From Fig.~\ref{fig:gain2} show the gains of MMALFM model over other competitors on the ``Clothing" and ``Yelp" datasets, respectively. Table~\ref{tab:coldstart} shows the corresponding percentage of cold-start users with 1 to 10 training samples in the two datasets.  Compared to BMF method which only uses ratings, substantially improvement has been achieved by TALFM. Moreover, TALFM greatly outperforms all the other baselines which also utilize reviews. Besides, we can see that the performance of MMALFM is comparable to that of TALFM in the cold-start setting. It indicates that the additional consideration of image information cannot further improve the performance in the cold-start setting

\subsection{Model Interpretability (\textbf{RQ5})} \label{sec:interpret}
In our model, a user's preference on an item is decomposed into user's preference on different aspects and the importance of those aspects. An aspect is represented as a distribution of latent topics discovered based on reviews. A user's attitude/sentiment on an aspect of the targeted item is decided by the latent factors (learned from ratings) associating with the aspect. Based on the topic distribution of an aspect ($\bm{\theta_{u,a}}$) and the word distribution of topics ($\bm{\phi_{t}}$), we can semantically represent an aspect by the top words in this aspect.  Specifically, the probability of a term $t$ in an aspect $a$ of a user $u$ can be  computed as $\sum_{k=1}^K\theta_{u,a,k}\phi_{k,t}$. The top 10 aspect terms of ``user\_2397", ``item\_137", and ``item\_673"  from the dataset discovered by our model are shown in Table~\ref{tab:aspects}. Notice that we removed the “background” words that belong to more than 3 aspects, because those words are not helpful on distinguishing different aspects. Examples of background words in the Yelp dataset include \emph{“nice”, “great”, “favor”, “love”, “amazing”, "bit", “pretty”, “well”, “back”}, etc. We can see that the found top terms highly match the corresponding aspects.

\begin{table}[]
	\small
	\centering
	\caption{Top ten words of five aspects for a user (index 2397) and two items (index 137 and 673) from the dataset.}
	\begin{tabular}{c|l|l} \hline
\multirow{5}{*}{User\_2397}	&	Food	&	sauce, fried, bread, fresh, huge, flavor, shrimp, dessert, dish	\\
	&	Ambience	&	nice, bar, atmosphere, location, friendly, inside, decor, staff, music	\\
	&	Price	&	expensive, high, cheap, pricey, decent, pay, reasonable, priced, deal	\\
	&	Service	&	table, server, friendly, minutes, nice, staff, asked, make, seated	\\
	&	Misc.	&	never, give, restaurant, times, stars, friends, night, places, dinner	\\  \hline
\multirow{5}{*}{Item\_137}	&	Food	&	sauce, salad, fries, dish, cheese, dishes, burger, fresh, crab	\\
	&	Ambience	&	bar, atmosphere, patio, area, inside, wine, small, cool, decor	\\
	&	Price	&	price, worth, prices, better, bit, meal, sauce, dishes, quality	\\
	&	Service	&	table, bar, friendly, wait, server, staff, minutes, beer, atmosphere	\\
	&	Misc.	&	eat, dinner, vegas, experience, wait, friends, times, never, give	\\  \hline
\multirow{5}{*}{Item\_673}	&	Food	&	nigiri, sake, tempura, shrimp, sauce, items, poke, crab, chef	\\
	&	Ambience	&	atmosphere, friendly, bar, staff, inside, area, spot, monta, feel	\\
	&	Price	&	price, worth, prices, nigiri, sake, tempura, items, lunch, special	\\
	&	Service	&	service, table, server, friendly, minutes, staff, nice, asked, seated	\\
	&	Misc.	&	restaurant, times, give, favorite, night, places, stars, friends, vegas	\\  \hline

	\end{tabular}
	\label{tab:aspects}
\end{table}

\begin{table}[]
	\small
	\centering
	\caption{Interpretation for why the ``user 2397" rated ``item 137" and ``item 673" with 5 and 2, respectively.}
	\begin{tabular}{c|l|ccccc} \hline
Item	&	Aspect	&	Food	&	Ambience	&	Price	&	Service	&	Misc.	\\ \hline
\multirow{3}{*}{Item\_137}	&	Importance	&	0.3815	&	0.1034	&	0.0723	&	0.2038	&	0.2390	\\
	&	Matching	&	0.5672	&	0.4523	&	0.5329	&	0.6021	&	0.7138	\\
	&	Polarity	&	\textbf{\color{red}+}	&	\textbf{\color{red}+}	&	\textbf{\color{blue}-}	&	\textbf{\color{red}+}	&	\textbf{\color{red}+}	\\ \hline
\multirow{3}{*}{Item\_673}	&	Importance	&	0.3726	&	0.0794	&	0.0853	&	0.2076	&	0.2551	\\
	&	Matching	&	0.1813	&	0.6535	&	0.4512	&	0.6018	&	0.7093	\\
	&	Polarity	&	\textbf{\color{blue}-}	&	\textbf{\color{blue}-}	&	\textbf{\color{red}+}	&	\textbf{\color{red}+}	&	\textbf{\color{blue}-}	\\ \hline
	\end{tabular}
	\label{tab:explaination}
\end{table}
 Next, we illustrate the interpretability of our model on high or low ratings by examples from the dataset. Table~\ref{tab:explaination} shows the aspect importance (i.e., $\rho_{u,i,a}$ in Eq.~\ref{eq:rho}) of the ``user\_2397",  the aspect matching scores (i.e., $s_{u,i,a}$ in Eq.~\ref{eq:jsd}) as well as sentiment polarity (obtained by Eq.~\ref{eq:re2}) on the five aspects with respect to ``item\_137" and ``item\_673" . From the results, we can see that the user cares the aspects of ``Food", ``Service", and ``Ambience", while pays less attention to ``Price".\footnote{``Others" considers all the factors besides the four aspects, we do not discuss here.} We can see that the properties of ``Item\_137" match ``user\_2397"'s preference on ``Food" and ``Service" well, and the user also has a positive sentiment on these aspects. While for ``Item\_673", the ``Food" aspect does not match user's preference well, which can also be observed from the top terms in Table~\ref{tab:aspects}; it has a good match on ``Ambience", while the user has a negative sentiment on this aspect. As a reminder, the aspect matching is based on the reviews. It is possible that ``Item\_673" contains many comments on aspect ``Ambience", but those comments are from the negative perspective. Overall, \emph{aspect importance}, \emph{aspect matching score}, and \emph{sentiment polarity} in Table~\ref{tab:explaination} could explicitly explain the reasons why the user give a high rating (i.e., 5) to ``Item\_137" while a low rating (i.e., ``2") to ``Item\_673". From the examples, we can see that our model could provide explanations for the recommendations in depth.

\subsection{Top-n Recommendation (\textbf{RQ6})} \label{sec:topn}
Top-n recommendation aims to recommend a set of $n$ top-ranked items that will be of interest to a certain user. Compared with rating prediction, top-n recommendation is a more practical task in real commercial systems because they expect the recommendations to customers could be converted into purchase behaviours.  To achieve good performance on this task, it is crucial to accurately capture a user's preference on each item. For a specific item, our model infers a user's preferences on this item by considering how does this item match this user's preferences on different aspects and the importance of those aspects. For the top-n task, our model can rank the items according to the predicted ratings and so as to generate the recommended ranking list.  In this section, we evaluate the performance of our model on the top-n recommendation task.

\subsubsection{Experimental Setup}
In this experiment, we compare our model with the following baselines: BPR-MF~\cite{rendle2009bpr} BPR-HFT~\cite{mcauley2013hidden}, VBPR~\cite{he2016vbpr},  EFM~\cite{zhang2014explicit}, and TALFM. Among those competitors: BPR-MF is designed for the top-n task by using the pair-wise learning to rank strategy; BPR-HFT exploits both rating and review information in preference modeling and applies the BPR strategy; VBPR is a visual-based BPR algorithm; EFM extracts product aspects and sentiment analysis for recommendation; and finally, TALFM is a variant of our model by excluding the visual information.   As EFM and TALFM have been described in Section~\ref{sec:comp}, we briefly introduce the other three methods.
\begin{itemize} [align=left,style=nextline,leftmargin=*,labelsep=\parindent,font=\normalfont]
    \item \textbf{BPR~\cite{rendle2009bpr}.} Bayesian Personalized Ranking (BPR) combines the matrix factorization method with a pair-wise learning to rank loss function.  And it has been proven to be is a competitive baseline for this task~\cite{he2016fast,he2017neural,zhang2017joint}. Notice that in this method, only the rating information is used.
    \item \textbf{BPR-HFT~\cite{mcauley2013hidden}.} The original HFT model is designed for rating prediction method. BPR-HFT extends the HFT for the top-n recommendation task by adding a BPR loss on top of HFT. Therefore, BPR-HFT leverages both rating and review information.
    \item \textbf{VBPR~\cite{he2016vbpr}.} This is a Visual Bayesian Personalized Ranking method for top-n recommendation, which is the state-of-the-art method for recommendation based on visual images of the products.
\end{itemize}

\begin{table}[]
	\centering
	\caption{Performance of Top-$n$ recommendation in different datasets. }
	\begin{tabular}{|l|c|cccccc|} \hline
Dataset	&	Metric	&	BPR	&	BRP-HFT	&	VBPR	&	EFM	&	TAFLM	&	MMALFM	\\ \hline \hline
\multirow{3}{*}{Beauty}	&	HR	&	8.241	&	8.268	&	5.961	&	9.312	&	10.15	&	11.54	\\
	&	Precision	&	1.143	&	1.132	&	0.902	&	1.293	&	1.203	&	1.216	\\
	&	NDCG	&	2.753	&	2.934	&	1.901	&	3.478	&	3.373	&	3.654	\\
\multirow{3}{*}{CDs}	&	HR	&	8.554	&	9.926	&	2.933	&	8.954	&	11.38	&	13.53	\\
	&	Precision	&	1.085	&	1.268	&	0.328	&	1.071	&	1.476	&	1.523	\\
	&	NDCG	&	2.009	&	2.661	&	0.631	&	2.936	&	3.853	&	4.112	\\
\multirow{3}{*}{Cell Phone}	&	HR	&	5.273	&	8.125	&	5.002	&	7.525	&	10.86	&	11.76	\\
	&	Precision	&	0.595	&	0.858	&	0.507	&	0.819	&	1.087	&	1.103	\\
	&	NDCG	&	1.998	&	3.151	&	1.797	&	3.193	&	4.236	&	4.511	\\
\multirow{3}{*}{Clothing}	&	HR	&	1.767	&	2.872	&	1.557	&	2.627	&	3.156	&	3.873	\\
	&	Precision	&	0.185	&	0.297	&	0.166	&	0.284	&	0.385	&	0.404	\\
	&	NDCG	&	0.601	&	1.067	&	0.56	&	1.076	&	1.682	&	1.799	\\
\multirow{3}{*}{Movies}	&	HR	&	4.421	&	6.378	&	2.976	&	5.368	&	8.934	&	11.87	\\
	&	Precision	&	0.528	&	0.776	&	0.324	&	0.575	&	0.976	&	1.232	\\
	&	NDCG	&	1.267	&	2.092	&	0.849	&	2.031	&	3.523	&	3.907	\\
\multirow{3}{*}{Yelp}	&	HR	&	23.52	&	30.45	&	12.35	&	28.84	&	38.56	&	40.35	\\
	&	Precision	&	2.162	&	4.228	&	1.328	&	3.746	&	5.762	&	6.171	\\
	&	NDCG	&	5.234	&	8.127	&	4.305	&	7.329	&	9.674	&	10.13	\\ \hline

	\end{tabular}
	\label{tab:topnresult}
\end{table}

 For each dataset, 70\% of each user is used for training, and the remain 30\% is used for testing. Notice that for the five Amazon product datasets, we used the exactly same training and testing split as the experiments in~\cite{zhang2017joint}.\footnote{\emph{The exactly same split} means that for each user, which samples are used in training and which samples are used in testing are exactly the same.} Thus, the results of some competitors in Table~\ref{tab:topnresult} are directly cited from~\cite{zhang2017joint}. The following three metrics are used in the evaluation:
 \begin{itemize} [align=left,style=nextline,leftmargin=*,labelsep=\parindent,font=\normalfont]
    \item \textbf{Precision}: it is the percentage of correctly recommended items (namely, the items that purchased by the targeted user) in a user's recommendation list.
    \item \textbf{Hit Ratio (HR)}: it represents the percentage of users that have at least one correctly recommended item in their lists. It evaluates how likely the recommendation system will provide at least one good recommendation to different users.
    \item \textbf{NDCG}: this measure takes the positions of correctly recommended items into considerations. As users usually only focus on the top few results in a recommendation list, it is important to rank the correct ones at the top positions.
 \end{itemize}
For each evaluation metric, the performance are evaluated based on the top 10 results; and we report the average value across all the testing users.

\subsubsection{Experimental Results}
The experimental results of all the considered competitors across different datasets are reported in Table~\ref{tab:topnresult}. Notice that the values shown in this table do not reflect the performance of those methods in real scenarios, because the measures are calculated based on the limited interactions of users in the datasets. It is possible that the recommended items are actually liked by the target user but this user has not noticed them.

From the results, we can observe that our model MMALFM, which leverages images, review and rating information in preference modeling, achieves the best performance in terms of different evaluation metrics across different datasets. The results demonstrate the potential of integrating different information sources in recommendation and the effectiveness of our model on integrating those information sources for user modeling. The models which utilize both review and rating information (i.e., BPR-HFT, EFM, TAFLM) greatly outperform the BPR model. It is expected and consistent with the observations in the rating prediction task. Because textual reviews contains rich information about user preferences and item characteristics, the utilization of such information in preference modeling can infer users' preferences on items more accurately and thus achieve higher recommendation accuracy. It is worth mentioning that TALFM outperforms BPR-HFT and EFM by a large margin, which also demonstrates the benefits of the fine-grained modeling of users' preferences on different aspects.

The integration of images benefits user preference modeling in our model, which can be observed by comparing the results of MMALFM with that of TALFM. Remind that the only difference between MMALFM and TALFM is that the former takes advantage of item images into the modeling of user preferences and item characteristics on different aspects. The better performance of MMALFM demonstrates that item visual appearances indeed affect user preference and can be exploited to improve recommendation performance, especially for the item of which visual appearance is an important aspect, such as clothing. Although item images are useful, the performance of using image alone is limited, which can be observed from the performance of VBPR. The performance of VBPR is not as good as BPR, which is based on the collaborative filtering mechanism without utilizing any content information. Besides, for the items of which are difficult to make judgments based on their appearance, the benefit from images in addition to reviews becomes limited, such as ``Movies" and ``CDs".

From the results,  we can safely conclude that (1) modeling user preference on different aspects has the potential to improve recommendation accuracy; (2) taking item images into recommender systems could infer user preference on item more accurately and thus achieve better performance; (2) our model could effectively capture user preferences on different aspects of items by integrating user reviews and item images.\footnote{Notice that the objective function of MMALFM is to minimize the rating prediction error, which is not particularly designed for ranking. Although its performance is better than the consideration competitors, it is attributed to its powerful mechanism on preference modeling (comparing to BPR-HFT and EFM) and the integration of images. In fact, its performance is not as good as JRL~\cite{zhang2017joint}, which uses deep learning techniques to integrate rating, review and image information for recommendation. JRL is a carefully designed model for the top-n recommendation by using a pair-wise learning to ranking objective function in training. Therefore, it is expected that MMALFM cannot compete JRL on the top-n recommendation task. As a reminder, the goal of this paper is to verify the effectiveness of modeling user preference on different aspects and the usefulness of image features. It is worth mentioning that it is an interesting future work to develop a MMALFM model for top-n recommendation.}

\section{Conclusions} \label{sec:concl}
In this paper, we present a multi-modal aspect-aware latent factor model for rating prediction and investigate the utility of item images on the performance. Based on user reviews and item images, a multi-modal aspect-aware topic model (MATM) is developed to learn users' interests and items' properties. Furthermore, an aspect-aware latent factor model (ALFM) is proposed to learn
aspect-aware latent factors by integrating results from MATM. The proposed model learns both aspects ratings and aspect importance to predict the overall ratings. Comparing to existing  review-based personalized rating prediction methods, our model has the advantages of learning the interaction between latent topics and latent factors on the semantic ``aspect" level. Experiments on a public accessible dataset demonstrate the superiority of our model, especially for users who have few ratings. The results also show that item images with visual features, which are related to important item properties, can improve the performance to some extent. Furthermore, our model could interpret the recommendation results in great detail.

\section*{acknowledgements}
This research is supported in part by the National Research Foundation, Prime Minister’s Office, Singapore under its International Research Centre in Singapore Funding Initiative;  the National Natural Science Foundation of China (Grant No. 61603233); and the Shaanxi Natural Science Foundation Youth Project (Grant No. 2017JQ6076).

\bibliographystyle{ACM-Reference-Format}
\bibliography{mmalfm}

\end{document}